\documentclass[preprint,12pt,3p,fleqn]{elsarticle}
\pdfoutput=1
\usepackage{amssymb}
\usepackage{amsthm}
\usepackage{amsmath}
\usepackage{multirow}
\usepackage{booktabs}
\usepackage{graphicx}
\usepackage{float}
\usepackage{url}
\usepackage{epstopdf}
\usepackage{pdflscape}
\usepackage{amsthm}
\usepackage{amsmath}
\usepackage[english]{babel}
\usepackage{subfigure}
\usepackage{threeparttable}

\usepackage{algorithm}
\usepackage{algorithmic}

\usepackage{xcolor}

\usepackage{tabularx}
\usepackage{graphicx}
\usepackage{caption}

\usepackage{threeparttable}
\usepackage[figuresright]{rotating}
\usepackage[toc,page,title,titletoc,header]{appendix}

\makeatletter

\newcommand{\Rmnum}[1]{\expandafter\@slowromancap\romannumeral #1@}
\makeatother

\journal{XXX}

\begin{document}
\captionsetup{font={scriptsize}}
\captionsetup[figure]{labelformat={default},labelsep=period,name={Fig.}}

\begin{frontmatter}
	\title{Multi-level adaptive particle refinement method with large refinement scale ratio and new free-surface detection algorithm for complex fluid-structure interaction problems}
	
	\author[a]{Tianrun Gao}
    \author[a]{Huihe Qiu}
	\author[a,b,c]{Lin Fu\corref{cor1}}
	\ead{linfu@ust.hk}
	\cortext[cor1]{Corresponding author.}

	\address[a]{Department of Mechanical and Aerospace Engineering, The Hong Kong University of Science and Technology, Clear Water Bay, Kowloon, Hong Kong}
	\address[b]{Department of Mathematics, The Hong Kong University of Science and Technology, Clear Water Bay, Kowloon, Hong Kong}
	\address[c]{Shenzhen Research Institute, The Hong Kong University of Science and Technology, Shenzhen, China}
	
	\begin{abstract}
		Fluid-Structure Interaction (FSI) is a crucial problem in ocean engineering. The smoothed particle hydrodynamics (SPH) method has been employed recently for FSI problems in light of its Lagrangian nature and its advantage in handling multi-physics problems. The efficiency of SPH can be greatly improved with the Adaptive Particle Refinement (APR) method, which refines particles in the regions of interest while deploying coarse particles in the left areas. In this study, the APR method is further improved by developing several new algorithms. Firstly, a new particle refinement strategy with the refinement scale ratio of 4 is employed for multi-level resolutions, and this dramatically decreases the computational costs compared to the standard APR method. Secondly, the regularized transition sub-zone is deployed to render an isotropic particle distribution, which makes the solutions between the refinement zone and the non-refinement zone smoother and consequently results in a more accurate prediction. Thirdly, for complex FSI problems with free surface, a new free-surface detection method based on the Voronoi diagram is proposed, and the performance is validated in comparison to the conventional method. The improved APR method is then applied to a set of challenging FSI cases. Numerical simulations demonstrate that the results from the refinement with scale ratio 4 are consistent with other studies and experimental data, and also agree well with those employing the refinement scale ratio 2. A significant reduction in the computational time is observed for all the considered cases. Overall, the improved APR method with a large refinement scale ratio and the new free-surface detection strategy shows great potential in simulating complex FSI problems efficiently and accurately. 
		
	\end{abstract}

	
\end{frontmatter}

\section{Introduction}

FSI problems are commonly seen in bridge engineering, wind engineering, and ocean engineering, e.g., Vortex-Induced-Vibration (VIV) of the bridges or high buildings, wing-flapping of birds, and swimming behaviors of fish, etc. A lot of FSI problems involve large deformation of structures and complex behaviors of violent flows. These FSI features pose a great challenge to high-fidelity numerical simulations. The traditional numerical strategy for FSI usually deploys the Finite Element Method (FEM) in the structure domain and the standard Eulerian mesh-based method or the Arbitary-Lagrangian-Eulerian (ALE) method in the fluid domain  \cite{takashi1994ale}. A major difficulty of these conventional methods for the FSI simulation is the mesh compatibility between the fluid and the structure surface. The Immersed Boundary Method (IBM) \cite{peskin2002immersed} is an alternative for FSI problem, where a background Eulerian mesh is deployed in the fluid domain while the Lagrangian mesh is employed for the structure. This strategy generally leads to accuracy degeneration at the interface due to the mismatch of the two different meshes. 

On the other hand, the meshless particle method is based on the Lagrangian concept, and is suitable for simulating violent flows with free-surface flows and large deformation of structures. To date, there have been many particle-based methods, e.g., SPH methods \cite{sun2019study}\cite{zhang2021multi}\cite{o2021fluid}\cite{liu2013numerical}, Moving Particle Semi-implicit (MPS) methods \cite{khayyer2019multi}\cite{khayyer20213d}, Finite Particle Methods (FPM) \cite{zhang2019predicting}\cite{idelsohn2008unified}, etc. Among them, SPH is an extremely popular method, which is initially proposed to solve problems related to astrophysics. Owing to the Lagrangian nature, SPH is inherently suitable for the FSI problems with large deformation and violent flows. Over the past decades, extensive studies with SPH methods have been reported on diverse problems, e.g, violent flows \cite{colagrossi2003numerical}, structure mechanics \cite{ganzenmuller2015hourglass}\cite{zhu2022dynamic}, FSI analyses \cite{liu2013numerical}\cite{sun2021accurate}, two-phase flows \cite{colagrossi2003numerical}\cite{ji2019new}\cite{ji2019lagrangian}, and bio-applications \cite{zhang2021integrative}\cite{jacob2021arbitrary}\cite{lai2022multiphase}, etc. In a word, SPH has become a promising method for simulating complex flows and their interactions with structures. 

In most circumstances, the scale of structures is small compared to the whole computational domain, and the particle spacing is even smaller in the structure domain. The computational cost is extremely high if this particle spacing is used across the entire computational domain. As a result, refining the particles in the area of interest while maintaining a coarse resolution in the left areas is a computationally cheap method. To this end, a set of multi-resolution methods for SPH has been developed so far. The particle splitting approach is initially devised by Feldman and Bonet \cite{feldman2007dynamic}, where the coarse particles split into finer particles and the splitting is irreversible. To conquer the drawback of the splitting method, Vacondio et al. \cite{vacondio2013variable} propose the particle merging approach, and incorporate these two methods for SPH. Based on the idea of particle splitting and merging, many improved methods have been proposed \cite{kitsionas2002smoothed}\cite{reyes2013dynamic}. In these methods, the particles of different resolutions can interact with each other, which may lead to the accuracy degeneration; moreover, the frequent merging of particles usually requires complicated implementations. To avoid the merging technique, Barcarolo et al.  \cite{barcarolo2014adaptive} propose an Adaptive Particle Refinement (APR) method, where coarse particles are retained after splitting and wait for re-activation. To avoid the interaction of particles across different resolutions, Chiron et al. \cite{chiron2018analysis} extend the method of Barcarolo et al. \cite{barcarolo2014adaptive} by adding a so-called guard zone which works as the data exchange channel across different resolutions, restricting the particles to interact only with those in the same resolution. Though APR achieves great success for multi-resolution SPH methods, there are still some problems. Firstly, the `inactive' particles, which do not participate in the physical time evolution of SPH, usually do not have an isotropic particle distribution in the so-called guard zone, and this may lead to poor simulation accuracy in this area. Secondly, the `inactive' coarse particles are not removed from the computational domain after the new particle generation, and as such, the particle density is very high in the finest particle level, which increases both the memory loads and the number of particle neighbors, and finally increases the overall computational costs. There has been a set of studies \cite{sun2019extension}\cite{hermange20193d}\cite{lyu20223d}\cite{gao2022block} on FSI problems with the APR strategy, where the refinement scale ratio 2 is typically used. A small scale ratio of 2 usually requires too many redundant resolution levels to achieve the targeted finest resolution, which takes extra storage memory and computational cost. Furthermore, the data communications between many different resolution levels may degenerate the solution accuracy. Hence, a larger refinement scale ratio 4 with multiple resolution levels is expected to decrease the redundant particle number and save computational costs effectively. However, to the authors' knowledge, the large refinement scale ratio of 4 in APR with multiple resolution levels has never been reported and verified.

Usually, the neighbor kernel support of free-surface particles is incomplete, and some physical boundary conditions or numerical methods are required to be implemented on these free-surface particles, e.g., the surface tension effect should be considered for the particles on the liquid surface when simulating microscale flows; convection conditions should be applied to surface particles to simulate heat convection in the ambient environment. Moreover, in order to ensure an isotropic particle distribution, the transport velocity method \cite{adami2013transport}\cite{oger2016sph} requires detecting the particles on the surface to enforce different particle shifting displacements according to their distance to the free surfaces. The widely used method is based on solving eigenvalues of a correction matrix, proposed by Marrone et al. \cite{marrone2010fast}. This method involves two steps and the physical meaning of using eigenvalues for free-surface detection is not straightforward. This detection method may make a mistake when there is a thin gap area between two surfaces. Developing an efficient and accurate method for free-surface detection is still a research topic. 

In this study, an improved robust APR method with multi-level resolutions, large refinement scale ratio, and regularized transition sub-zone is proposed. In order to reduce computational costs and unnecessary resolution levels, a large refinement scale ratio 4 with multiple levels is explored and verified. In order to solve the problem of the irregular particle distribution in the guard zone, the regularized transition sub-zone is deployed in the guard zone to render the particle distribution isotropic.  A corresponding multi-resolution data structure based on the multi-resolution cell-linked list method is incorporated for efficient neighbor searching. A new free-surface detection method based on the Voronoi diagram is proposed to facilitate the implementation of the particle shifting technique. With the improved APR method, FSI problems will be investigated covering a set of challenging problems, e.g., thin-structure FSI problem, two-phase FSI problem, water impact problem, and fishlike swimming problem. The remainder of this paper is arranged as follows. In section \ref{sect1}, the governing equations of fluids and solids, and the FSI coupling strategy are introduced. Moreover, the two-phase FSI formulations are also elaborated. In section \ref{sect2}, a novel free-surface detection method based on the Voronoi diagram will be proposed. In section \ref{sect3}, the improved APR method with large refinement scale ratio and regularized transition sub-zone is introduced. In section \ref{sect4}, a set of challenging cases will be simulated with the improved APR method, and the performance of the new free-surface detection method will be assessed. Finally, concluding remarks will be given in the last section. 

\section{FSI and two-phase model}
\label{sect1}
\subsection{Governing equations of the fluid phase} 
\subsubsection{Single-phase model} 
The discretized SPH formulations of the governing equations for the fluid phase are expressed as
\begin{equation}
	\label{eq:discr_sph1}
	\left\{
	\begin{array}{l}
		\begin{aligned}
			\vspace{1ex}
			&\dfrac{d {\rho}_{i}}{d t} ={\rho}_{i} \sum_{j}  \boldsymbol{v}_{i j} \cdot \boldsymbol{\nabla}_{i} W_{i j} V_{j}+\delta h c_{0} \sum_{j} \boldsymbol{\Phi}_{i j} \cdot \boldsymbol{\nabla}_{i} W_{i j} V_{j},\\
			&\dfrac{d\boldsymbol{v}_{i}}{d t} =  -\dfrac{1}{\rho_i} \sum_{j} \left(p_{i}+p_{j}\right) \boldsymbol{\nabla}_{i} W_{i j} V_{j} + \dfrac{1}{\rho_i}\left( \boldsymbol{F}_v + \boldsymbol{F}_{Iv} \right)  + \boldsymbol{f},\\
			&\dfrac{d\boldsymbol{r_{i}}}{dt} = \boldsymbol{\tilde{v}}_{i},\\
			&p_{i} =c^2_0 \left({\rho}_{i}-{\rho}_{0}  \right),
		\end{aligned}
	\end{array}
	\right.
\end{equation}
where $\rho$, $\boldsymbol{v}$, $p$, $\eta$, $\boldsymbol{f}$ and $\boldsymbol{r}$ denote the density, velocity, pressure, dynamic viscosity, body force and particle coordinates, respectively. Here, $ \boldsymbol{v}_{i j} $ and $ \boldsymbol{r}_{i j} $ are defined as $\boldsymbol{v}_{i j}=\boldsymbol{v}_{i} - \boldsymbol{v}_{j} $, $\boldsymbol{r}_{i j}=\boldsymbol{r}_{i} - \boldsymbol{r}_{j} $, and $V_{j}=m_{j}/\rho_{j}$ is the volume of the particle. $\nabla_{i} W_{i j}$ denotes the gradient of the kernel function $W(\left\| \boldsymbol{r}_{i j}\right\|,h)$. The Gaussian kernel is employed with a smoothing length of $h=1.2\Delta {x}$, where $\Delta {x}$ is the initial particle spacing, and the cut-off parameter $q$ is set as 3, \textcolor{black}{then the kernel radius is $qh=3.6\Delta {x}$}. $ \boldsymbol{\tilde{v}}_{i} $ is the transport velocity used to regularize the particle distribution, with $  \boldsymbol{\tilde{v}}_{i}= \boldsymbol{{v}}_{i} + \delta \boldsymbol{v}_i $. 
The last equation relates the pressure and density of particle $ i $, where $ \rho_0 $ is the initial density of the particle, and $ c_0 $ is the artificial sound speed, which satisfies $c_{0} \geq 10\left( U_{max}, \sqrt{\frac{p_{max}}{\rho_0}} \right)$ \cite{sun2017deltaplus} to ensure the weakly-compressible condition, and $U_{max}$ and $p_{max}$ are the expected maximum velocity and pressure in the computational domain. 

A simplified dissipation term $ \boldsymbol{\Phi}_{i j} $ \cite{molteni2009simple} in the density equation is used to alleviate pressure fluctuation, and can be expressed as
\begin{equation}
	\boldsymbol{\Phi}_{i j} =2\left( \rho_{i}- \rho_{j} \right) \dfrac{ \boldsymbol{r}_{i j} }{\left\| \boldsymbol{r}_{i j}\right\| ^{2} }.
\end{equation}
For viscous flows, the physical viscosity term is written as
\begin{equation}
	\boldsymbol{F}_v= \sum_{j}\frac{\left(\eta_{i}+\eta_{j}\right) \boldsymbol{r}_{i j} \cdot \boldsymbol{\nabla}_{i} W_{i j}V_{j}}{\left(\left\| \boldsymbol{r}_{i j}\right\| ^{2}+ 0.01h^{2}_{i}   \right)} \boldsymbol{v}_{i j},
	\label{visco}
\end{equation}

while for inviscid flows, an artificial viscosity term is usually used instead, i.e.,
\begin{equation}
	\boldsymbol{F}_{Iv}=\alpha \rho_i h_i c_{0} \sum_{j}  \boldsymbol{v}_{ij} \cdot \dfrac{ \boldsymbol{r}_{i j} }{\left\| \boldsymbol{r}_{i j}\right\| ^{2} }  \boldsymbol{\nabla}_{i} W_{i j} V_{j}.
\end{equation}
Here, the two coefficients are chosen as $ \delta=0.1 $ \cite{molteni2009simple}\cite{marrone2011delta} and $\alpha=0.02$ \cite{molteni2009simple}, if not explicitly specified otherwise. 

To ensure an isotropic particle distribution, the shifting velocity $\delta \boldsymbol{v}_i $ \cite{antuono2021delta} is employed, which is expressed as
\begin{equation}
	\label{eq:background_P}
	\left\{
	\begin{array}{l}
		\begin{aligned}
			&\delta \boldsymbol{v}_i=\min\left(\left\| \delta \boldsymbol{v}_i^{\prime}\right\|, U_{max}/2 \right) \frac{\delta \boldsymbol{v}_i^{\prime}}{\left\| \delta \boldsymbol{v}_i^{\prime}\right\|},\\
			&\delta \boldsymbol{v}_i^{\prime} = -{\rm Ma}(2h) c_{0} \sum_{j} \left[1+0.2\left(\dfrac{W_{i j}}{W(\Delta x,h)}\right)^{4}  \right] \boldsymbol{\nabla}_{i} W_{i j} V_{j}, 
		\end{aligned}
	\end{array}
	\right.
\end{equation}
where ${\rm Ma}=U_{max}/c_{0}$. Referring to Sun et al. \cite{sun2017deltaplus}, the shifting induced by  Eq. (\ref{eq:background_P}) is much smaller than the particle spacing in each time step. Therefore, the correction of field values using first-order Taylor expansion  after the particle shifting is not needed \cite{xu2009accuracy}. 
It is worth noting that the transport velocity $ \boldsymbol{\tilde{v}}$ of the  particles in the surface area is set equal to their real velocity, considering their incomplete kernel support. Herein, the particles in the `surface area' means the particles within the radius $ 4\Delta x $ of the `surface particles'. The `surface particles' are identified with a free-surface detection method. In this study, a new detection method of free-surface particles will be presented in section \ref{sect2}.

The so-called Kick-Drift-Kick scheme \cite{monaghan2005smoothed}\cite{adami2013transport}\cite{zhang2017generalized} is used for the time integration, written as
\begin{equation}
	\label{eq:scheme1}
	\begin{array}{l}
		\begin{aligned}
			&\boldsymbol{v}^{n+\frac{1}{2}} = \boldsymbol{v}^{n} + \frac{\Delta t_{F}}{2} {\left( \dfrac{d \boldsymbol{v}}{d t} \right)}^{n},\\
			&\boldsymbol{\tilde{v}}^{n+\frac{1}{2}} = \boldsymbol{v}^{n+\frac{1}{2}} + \delta \boldsymbol{v}_i^n,\\
			&\boldsymbol{x}^{n+1} = \boldsymbol{x}^{n} + {\Delta t_{F}} \boldsymbol{\tilde{v}}^{n+\frac{1}{2}},\\
			&\rho^{n+1} = \rho^{n} + \Delta t_{F} {\left(  \dfrac{d\rho}{d t} \right)}^{n+\frac{1}{2}},\\
			&\boldsymbol{v}^{n+1} = \boldsymbol{v}^{n+\frac{1}{2}} + \frac{\Delta t_{F}}{2} {\left( \dfrac{d \boldsymbol{v}}{d t} \right)}^{n+1},\\
		\end{aligned}
	\end{array}
\end{equation}
where the variables are calculated in sequence. The second equation in Eq. (\ref{eq:scheme1}) gives the relationship between the transport velocity $ \boldsymbol{\tilde{v}}$ and the shifting velocity $\delta \boldsymbol{v}_i $, and the third equation shows that the positions of particles are updated by the transport velocity $ \boldsymbol{\tilde{v}}$. The time step $ \Delta t_{F} $ of the integration is defined as
\begin{equation}
	\begin{aligned}
		&\Delta t_{a} = 0.25 \sqrt{\dfrac{h}{\left\|  \boldsymbol{a} \right\|  _{max}}}, \text{ } \Delta t_{v} = 0.25 \dfrac{h}{c_{0}+{\left\|  \boldsymbol{v} \right\| _{max}}},\\
		&\Delta t_{F} = \min(\Delta t_{a}, \Delta t_{v}).\\
	\end{aligned}
\end{equation}
\subsubsection{Two-phase model} 
When there are two fluid phases in the computational domain, the two phases will interact with each other. Referring to the model in \cite{he2022stable}\cite{sun2019study}, the governing equations can be expressed as
\begin{equation}
	\label{eq:discr_sph2}
	\left\{
	\begin{array}{l}
		\begin{aligned}
			\vspace{1ex}
			&\dfrac{d{\rho}_{i}}{d t} ={\rho}_{i} \sum_{j \in {\chi}_{i}}  \boldsymbol{v}_{i j} \cdot \boldsymbol{\nabla}_{i} W_{i j} V_{j}+\delta h c_{0} \sum_{j \in {\Omega}_{i}} \boldsymbol{\Phi}_{i j} \cdot \boldsymbol{\nabla}_{i} W_{i j} V_{j},\\
			&\dfrac{d \boldsymbol{v}_{i}}{d t} =  -\dfrac{1}{\rho_i} \sum_{j\in {\Omega}_{i}} \left(p_{i}+p_{j}\right) \boldsymbol{\nabla}_{i} W_{i j} V_{j} +\dfrac{1}{\rho_i} \left( \boldsymbol{F}_v^{j \in {\chi}_{i}} + \boldsymbol{F}_{Iv}^{j \in {\chi}_{i}}\right) + \dfrac{1}{\rho_i}\boldsymbol{F}^{interface}_{i} +  \boldsymbol{f},\\
			&\dfrac{d\boldsymbol{r_{i}}}{dt} = \boldsymbol{\tilde{v}}_{i},\\
			&p_{i} =c^2_0 \left({\rho}_{i}-{\rho}_{0}  \right)+p_{bg},
		\end{aligned}
	\end{array}
	\right.
\end{equation}
where, $ {\chi}_i $ denotes the particles having the same phase with particle $ i $, and $ {\Omega}_{i} $ denotes all the particles within the kernel support. In the continuity equation, the velocity term $\boldsymbol{v}_{i j} $ only comes from the same phase, which avoids directly using the velocity from the light phase, especially the velocity of light particles around the water jet. The same artificial sound speed is used for the two phases \cite{he2022stable}\cite{zhang2022efficient}. The term $\boldsymbol{F}^{interface}_{i}  $ \cite{grenier2009hamiltonian} in the momentum equation is employed to prevent phase penetration, expressed as
\begin{equation}
	\begin{aligned}
		\boldsymbol{F}^{interface}_{i} = -0.08\sum_{j\in \bar{\chi}_{i}} \left(\left\|p_{i} \right\| +\left\|p_{j} \right\| \right) \boldsymbol{\nabla}_{i} W_{i j} V_{j},
	\end{aligned}
\end{equation}
where $ \bar{\chi}_{i} $ denotes the particles having different phase with particle $ i $.  \textcolor{black}{For two-phase problems, the negative pressure may induce a numerical instability especially when the density ratio is large, and therefore a positive background pressure $ p_{bg} $ is used in this study \cite{chen2015sph}\cite{he2022stable}\cite{sun2019study}. According to the studies of  \cite{chen2015sph} and \cite{he2022stable}, the pressure will be set as zero when the particle pressure is negative to stabilize the simulation, and meanwhile, the corresponding density is modified as}
\begin{equation}
	\begin{aligned}
		&\rho_i = -\dfrac{p_{bg}}{c_0^2}+\rho_0, \text{  if  } p_i \le 0. 
	\end{aligned}
\end{equation}
Moreover, in order to further improve the numerical stability, the density re-initialization method  is used in the multiphase problem \textcolor{black}{\cite{chen2015sph}\cite{he2022stable}}.

\subsection{Governing equations of the solid phase} 
The discretized form of the governing equations for the solid dynamics can be expressed as 
\begin{equation}
	\label{eq:solid}
	\dfrac{d \boldsymbol{v}_{a}}{d t} =  -\dfrac{1}{\rho_{S}} \sum_{b} \left(\boldsymbol{\rm P}_{a} \boldsymbol{\rm L}_{0a}^{-1} + \boldsymbol{\rm P}_{b} \boldsymbol{\rm L}_{0b}^{-1} \right) \nabla_{0a} W_{0ab} V_{b} + \boldsymbol{f}+ \textcolor{black}{\dfrac{1}{\rho_{S}V_{b}} \boldsymbol{f}_{a}^{H G}},
\end{equation}
which describes the acceleration of particle $ a $ and its interaction with the neighboring particle $ b $. Here, the operator $\boldsymbol{\nabla}_{0}$ is with regard to the initial coordinates $\boldsymbol X$, and ${\boldsymbol{\rm P}}$ denotes the first Piola–Kirchhoff stress tensor. $ \rho_{S} $ is the density of the solid. 
The correction matrix $ \boldsymbol{\rm L}_{0a} $ is given as 
\begin{equation}
	\boldsymbol{\rm L}_{0a} = \sum_{b} ( \boldsymbol{X}_{b} -\boldsymbol{X}_{a} ) \otimes \nabla_{0a} W_{0ab} V_{b}.
\end{equation}
The displacement gradient $\boldsymbol{\rm F} = \dfrac{d \boldsymbol{x} }{d \boldsymbol{X}}$ can be discretized as
\begin{equation}
	\boldsymbol{\rm F}_{a} = \sum_{b} ( \boldsymbol{x}_{ b} -\boldsymbol{ x}_{a} ) \otimes \boldsymbol{\rm L}_{0a}^{-1} \boldsymbol{\nabla}_{0a} W_{0ab} V_{b}.
\end{equation}
With the displacement gradient, the deformation tensor Green–Lagrange strain is derived as
\begin{equation}
	\boldsymbol{\rm E} = \dfrac{1}{2}( { {\rm \boldsymbol{\rm F}}_{a}^{\rm T}} { \boldsymbol{\rm F}_{a}} - \boldsymbol{\rm I} ).
\end{equation}
In this study, the Saint Venant-Kirchhoff constitutive model is employed for elastic solids while  Mooney-Rivlin model is employed for the elastomer. Regarding Saint Venant-Kirchhoff constitutive model, the second Piola–Kirchhoff stress tensor $\boldsymbol{\rm S}$ is obtained as
\begin{equation}
	\boldsymbol{\rm S} = \lambda tr(\boldsymbol{\rm E} ) \boldsymbol{\rm I} + 2 \mu \boldsymbol{\rm E},
\end{equation}
where $\lambda$ and $\mu$ are Lam\'e parameters. Regarding Mooney-Rivlin model, the strain energy density function is expressed as
\begin{equation}
	w = C_{10}(\bar{I}_1 -3)+ C_{01}(\bar{I}_2 -3)+ \frac{1}{D_1}(J-1)^{2},
	\label{mooney}
\end{equation}
where $ C_{10} $, $ C_{01} $ and $ D_1 $ are the model parameters. $ J $ is the third invariant of displacement gradient tensor $\boldsymbol{\rm F} $, and $ \bar{I}_1 $ and $ \bar{I}_2 $ are the first and second invariants of the reduced left Cauchy–Green deformation tensor $ J^{-2/3}\boldsymbol{\rm F}\boldsymbol{\rm F}^{T} $. Referring to \cite{basar2000nonlinear}, the second Piola–Kirchhoff stress tensor  is written as
\begin{equation}
	\boldsymbol{\rm S}  =\frac{\partial w}{\partial{\boldsymbol{\rm E}}}= \boldsymbol{\rm F}^{-1}\frac{\partial w}{\partial{\boldsymbol{\rm F}}}=2\boldsymbol{\rm F}^{-1}\frac{\partial w}{\partial ({\boldsymbol{\rm F}\boldsymbol{\rm F}^T})}\boldsymbol{\rm F}.
\end{equation}
Finally, the first Piola–Kirchhoff stress tensor is obtained with
\begin{equation}
	\boldsymbol{\rm P} = \boldsymbol{\rm F} \boldsymbol{\rm S}.
\end{equation}

\textcolor{black}{In Eq. (\ref{eq:solid}), the term $\boldsymbol{f}_{a}^{H G}$ is used to alleviate the so-called ``hourglass instability''. This term is proposed by Ganzenmüller et al. \cite{ganzenmuller2015hourglass}, and is written as
\begin{equation}
	\begin{array}{l}
		\begin{aligned}
    	&\boldsymbol{f}_{a}^{H G}=-\frac{1}{2} \alpha^{\prime} \sum_{b} \frac{V_{a} V_{b} W_{0 a b}}{\boldsymbol X_{ab}^{2}}\left(E \delta_{a b}^{a}+ E \delta_{b a}^{b}\right) \frac{\boldsymbol{x}_{a b}}{\left\|\boldsymbol{x}_{a b}\right\|},\\
    	&{\delta}_{ab}^a= \left(\rm \boldsymbol{\rm F}_a \boldsymbol X_{ab} -\boldsymbol x_{ab} \right)\cdot \frac{\boldsymbol{x}_{a b}}{\left\|\boldsymbol{x}_{a b}\right\|}. 
		\end{aligned}
	\end{array}	
	\label{eq:h_g}
\end{equation}
Here we choose the coefficient $ \alpha^{\prime}=0.2 $ in this study. }

The Kick-Drift-Kick (KDK) \cite{monaghan2005smoothed}\cite{adami2013transport}\cite{zhang2017generalized} scheme is employed for the time integration, i.e.,
\begin{equation}
	\label{eq:scheme}
	\begin{array}{l}
		\begin{aligned}
			&\boldsymbol{v}^{n+\frac{1}{2}} = \boldsymbol{v}^{n} + \frac{\Delta t_{S}}{2} {\left( \dfrac{d \boldsymbol{v}}{d t} \right)}^{n},\\
			&\boldsymbol{x}^{n+1} = \boldsymbol{x}^{n} + {\Delta t_{S}} \boldsymbol{v}^{n+\frac{1}{2}},\\
			&\boldsymbol{v}^{n+1} = \boldsymbol{v}^{n+\frac{1}{2}} + \frac{\Delta t_{S}}{2} {\left( \dfrac{d \boldsymbol{v}}{d t} \right)}^{n+1}.\\
		\end{aligned}
	\end{array}
\end{equation}
The time step $\Delta t_{S}$ for the solid is chosen as 
\begin{equation}
	\Delta t_{S} = 0.5h/c_{s},\text{ }c_{s}= \sqrt{\frac{E (1-\nu)}{\rho_{S}(1+\nu)(1-2\nu)}},
\end{equation}
where $c_{s}$, $E$ and $\nu$ denote the elastic wave speed, Young's modulus and Poisson's ratio, respectively.

\subsection{Interaction strategy between fluid and structure} 
In this study, the one-way coupling strategy is employed for the FSI coupling. Usually, the time step of the fluid phase $ \Delta t_{F} $ is much larger than that of the solid phase $ \Delta t_{S} $, which means that many time steps of the solid phase are advanced within one time step of the fluid phase. At the beginning of each time evolution of the fluid phase, the solid particles are fixed as the wall boundary and the fluid particles will advance with time step $ \Delta t_{F} $. Then the updated field values of the fluid phase, such as pressure and velocity, will act as the external force for the solid phase, and the solid particles will advance with many time steps within the time step $ \Delta t_{F} $. Readers are referred to \cite{sun2021accurate} for more details.

The fixed wall boundary method is utilized for the wall boundary particles, and the virtual fluid variables of these wall boundary particles are interpolated from the fluid phase, as proposed by Adami et al. \cite{adami2012generalized}. Thus, the momentum equation for the fluid particles is further modified as
\begin{equation}
	\label{eq:discr_couple_f}
	\begin{array}{l}
		\begin{aligned}
			
			\dfrac{d \boldsymbol{v}_{i}}{d t} &=  -\dfrac{1}{\rho_{i}} \sum_{j\in F({\Omega}_{i})} \left(p_{i}+p_{j}\right) \boldsymbol{\nabla}_{i} W_{i j} V_{j}  -\dfrac{1}{\rho_i} \sum_{j\in S } \left(p_{i}+p_{j}^{\prime}\right) \boldsymbol{\nabla}_{i} W_{i j} V_{j}  \\
			& + \dfrac{1}{\rho_i}\boldsymbol{F}_v^{j\in F({\chi}_{i})} +  \dfrac{1}{\rho_i}\boldsymbol{F}_v^{j\in S } +\dfrac{1}{\rho_i}\boldsymbol{F}_{Iv}^{j\in F({\chi}_{i})} + \boldsymbol{f},
			
		\end{aligned}
	\end{array}
\end{equation}
where $p_{j}^{\prime}$ denotes the virtual fluid pressure of the solid particle $ j $. $F$ and $S$ signify the particle sets of fluid phase, and the solid or boundary phase, respectively. $ j\in F({\Omega}_{i}) $ denotes  a neighboring particle $ j $, which is in the kernel support ${\Omega}_{i}  $ and belongs to fluid phase $ F $. The same rule is for the definition of $ j\in F({\chi}_{i}) $.

Similarly, the fluid particles exert pressure and viscous force on the solid particles. The momentum equation for the solid particles is given as
\begin{equation}
	\label{eq:discr_couple_s}
	\begin{array}{l}
		\begin{aligned}
			
			\dfrac{d \boldsymbol{v}_{a}}{d t} &= -\dfrac{1}{\rho_{S}} \sum_{b \in S} \left(\boldsymbol{\rm P}_{a} \boldsymbol{\rm L}_{0a}^{-1} + \boldsymbol{\rm P}_{b} \boldsymbol{\rm L}_{0b}^{-1} \right) \boldsymbol{\nabla}_{0a} W_{0ab} V_{b}\\
			& -\dfrac{1}{\rho_{S}} \sum_{b\in F({\Omega}_{i})} \left(p_{a}^{\prime}+p_{b}\right) \boldsymbol{\nabla}_{a} W_{ab} V_{b} +
			\dfrac{1}{\rho_S}\boldsymbol{F}_v^{b\in F({\Omega}_{i})} + \boldsymbol{f}+ \textcolor{black}{\dfrac{1}{\rho_{S}V_{b}} \boldsymbol{f}_{a}^{H G}}.
			
		\end{aligned}
	\end{array}
\end{equation}

\textcolor{black}{Regarding the interaction equations of Eq. (\ref{eq:discr_couple_f}) and (\ref{eq:discr_couple_s}), both the pressure term and the viscosity term are antisymmetric. Therefore, in terms of the particle interaction strategy, it is conservative. In terms of  the time integral scheme for physical evolutions, it is usually not conservative in practice. The time step for the fluid is usually larger than that of the solid, and therefore, one fluid time step evolution will contain  many solid time step evolutions, and in this process, the ghost pressure and velocity of the solid particles will change, while these values do not change in the fluid momentum Eq. (\ref{eq:discr_couple_f}). As such, the momentum between fluids and solids will not be fully conservative. Nevertheless, the effect of interaction non-conservation is not prominent in the weakly-compressible simulations and the present numerical strategy has been widely used in many studies \cite{sun2021accurate}\cite{o2021fluid}\cite{meng2022hydroelastic}. Therefore, Eq. (\ref{eq:discr_couple_f}) and (\ref{eq:discr_couple_s}) are still used in this work. }

\section{A novel free-surface detection method}
\label{sect2}
\subsection{Existing free-surface detection methods }
To date, there are various methods \cite{dilts2000moving}\cite{koshizuka1996moving}\cite{lee2008comparisons}\cite{marrone2010fast} for free-surface detection based on either algebraic or geometric algorithms. The most widely used method is proposed by Marrone et al. \cite{marrone2010fast} based on the eigenvalue concept. 
This method is composed of two steps. In the first step, the minimum eigenvalue $ \lambda $ of the matrix $ \boldsymbol{B}_i^{-1} $ is determined, where $ \boldsymbol{B}_i $ is a renormalization matrix written as
\begin{equation}
	\label{eq:renorm}
	\textcolor{black}{\boldsymbol{B}_i=\left[ \sum_{j} \nabla_{i} W_{ij} \otimes\left(\boldsymbol{x}_{j}-\boldsymbol{x}_{i}\right)  V_{j}\right]^{-1}.}
\end{equation}
If $ \lambda \le 0.2 $, then the particle is defined as the surface particle. The corresponding normal direction $ \boldsymbol{n}_i $ of the surface particle can be obtained by
\begin{equation}
	\boldsymbol{n}_i=\frac{\boldsymbol{v}_i}{\left|\boldsymbol{v}_i\right|}, \quad \textcolor{black}{\boldsymbol{v}_i=-\boldsymbol{B}_i  \sum_{j} \nabla_{i} W_{ij} V_{j}}.
\end{equation}
In the second step, a scan region is deployed along the normal direction $ \boldsymbol{n}_i $ for searching and checking if there is a particle in this region. If there is no particle, the targeted particle $ i $ will be identified as the surface particle, and vice versa. 

This method is a two-step algorithm and the physical meaning of using  eigenvalues for free-surface detection is not straightforward. In this work, we will propose a novel method for free-surface detection based on the Voronoi diagram, which has high computational efficiency. 

\subsection{A novel free-surface detection method based on Voronoi diagram}
\subsubsection{Voronoi diagram}
\begin{figure}[htbp]
	\centering
	\includegraphics[width=0.4\textwidth]{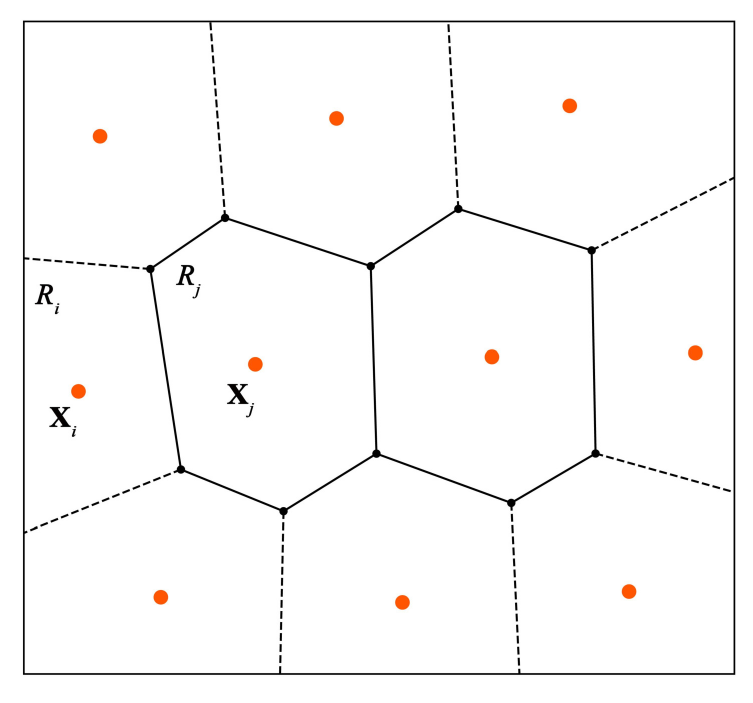}
	\caption{\textcolor{black}{Illustration of the Voronoi diagram constructed with 10 points.}}
	\label{voro_illu}
\end{figure}

Given a set of points (generators) $ \left\lbrace \mathbf{X}_i \right\rbrace $ with $ k $ elements in a  metric-space domain $ R $, a Voronoi diagram can be uniquely defined.  For the coordinate $ \mathbf{x} $ in the Voronoi cell $ R_{i} $ associated with the generator $ \mathbf{X}_i $, the distance between position $ \mathbf{x} $ and the corresponding generator $ \mathbf{X}_i $, i.e., $ \left\| \mathbf{x}- \mathbf{X}_i \right\|  $, is always not greater than the distance between position $ \mathbf{x} $ and the other generator $ \mathbf{X}_j $, i.e., $ \left\| \mathbf{x}- \mathbf{X}_j \right\|  $, and the Voronoi cell subdomain $ R_{i} $ can be mathematically defined as
\begin{equation}
	R_{i}=\left\{\mathbf{x} \in R \mid\left\|\mathbf{x}-\mathbf{X}_{i}\right\| \leqslant\left\|\mathbf{x}-\mathbf{X}_{j}\right\|, \forall j \neq i\right\},
\end{equation}
and for $j \neq i$, $ R_i \cap R_j =\emptyset$ and $ \cup_{i=1}^{k} R_{i}=R $. \textcolor{black}{For.pdf example, a Voronoi diagram can be constructed with 10 points as shown in Fig. \ref{voro_illu}. The entire domain $R$ is divided into 10 Voronoi cells $R_1$ $\sim$ $R_{10}$ separated with black lines. Each pair of points is symmetric with regard to the black lines. The solid lines are used to represent the finite lines, while the dashed lines are used for the infinite lines. For $\mathbf{X}_i$, the corresponding Voronoi cell $R_i$ contains infinite lines and therefore the cell is infinite, while for $\mathbf{X}_j$, the cell is finite. }

\subsubsection{Methodology of the new free-surface detection}
If the  positions $ \left\lbrace \boldsymbol{x}_{pi} \right\rbrace $  of all the SPH particles act as the Voronoi generators, a Voronoi diagram can be established. The centroid $ \boldsymbol{x}_{ci} $ of the Voronoi cell $ R_i $ can be calculated with 
\begin{equation}
	\boldsymbol{x}_{ci}=\dfrac{\int_{R_i} \mathbf{x} dS}{\int_{R_i} dS}.
\end{equation}
In 2D space, the Voronoi cell is a polygon, therefore the centroid $ \boldsymbol{x}_{ci} $ can be given as
\begin{equation}
	\label{eq:voro2}	
	\left\{
	\begin{array}{l}
		\begin{aligned}
			&\boldsymbol{x}_{ci}=\dfrac{1}{6A}\sum_{i=1}^{v} (\mathbf{v}_i + \mathbf{v}_{i+1})( v_i^x v^y_{i+1}-v^x_{i+1}v^y_i ),\\
			&A=\dfrac{1}{2}\sum_{i=1}^{v}( v_i^x v^y_{i+1}-v^x_{i+1}v^y_i ),
		\end{aligned}
	\end{array}
	\right.	
\end{equation}
where  $ \left\lbrace \mathbf{v}_{i} \right\rbrace $  denotes the coordinates of the polygon vertices, and $v_i^x$ and $v_i^y$ stand for the positions of $\mathbf{v}_{i}$ in $x$ and $y$ axes, respectively.
Usually, the centroid $ \boldsymbol{x}_{ci} $ does not coincide with the corresponding generator (particle) $ \boldsymbol{x}_{pi} $. If the two points coincide, the centroidal Voronoi tessellation is achieved, where the particle distribution is isotropic. Similarly, when the distribution of particles approaches isotropic, the centroid $ \boldsymbol{x}_{ci} $ and the corresponding generator $ \boldsymbol{x}_{pi} $ will be very close to each other and the distance between them should be at the magnitude of  $ \xi \Delta x $ ($ \Delta x $ denoting the initial particle spacing), i.e.,
\begin{equation}
	\label{eq:voro1}	
	\left\| \boldsymbol{x}_{pi}  - \boldsymbol{x}_{ci}\right\| \approx O(\xi \Delta x), 
\end{equation}
where $ \xi $ is a coefficient. The particles on the surface do not have a particle support beyond the geometry surface, therefore for those particles, Eq. (\ref{eq:voro1}) will not be satisfied. 

\begin{figure}[htbp]
	\centering
	\includegraphics[width=0.8\textwidth]{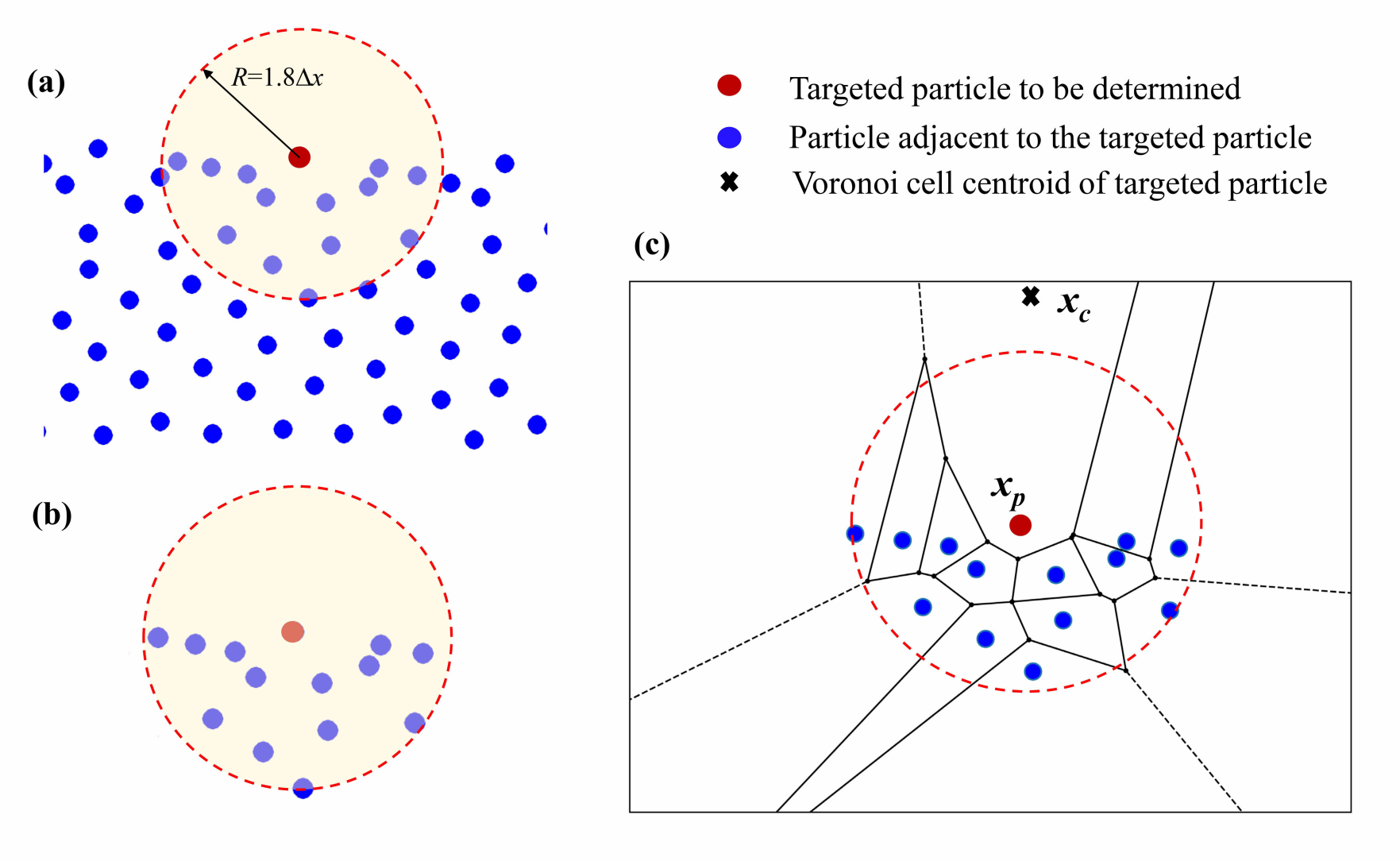}
	\caption{Schematic of the novel free-surface detection method: regarding the red particle to be determined, the neighboring particles (blue) within the radius of $ R=1.8\Delta x $ are found, shown in (a) and (b), then a Voronoi diagram is constructed with these particles, and the corresponding cell centroid $\boldsymbol{x}_c$ of the red particle $\boldsymbol{x}_p$ is calculated. If the condition $ \left\| \boldsymbol{x}_p  - \boldsymbol{x}_c\right\| \ge \xi\Delta x  $ is met or one edge of the Voronoi cell corresponding to the red particle is infinite, the targeted red particle will be determined as a surface particle.  }
	\label{voro_schm}
\end{figure}

Motivated by the analysis above, the surface particles can be detected based on the Voronoi diagram. For each particle, a Voronoi diagram can be constructed using its neighboring particles. The schematic of our proposed free-surface detection method with the Voronoi diagram is shown in Fig. \ref{voro_schm}. With regard to the red targeted particle in Fig. \ref{voro_schm} (a), the neighboring particles within the radius of $ R=1.8\Delta x $  are first identified and used as the Voronoi generators, as shown in Fig. \ref{voro_schm} (b). Then the corresponding Voronoi diagram is constructed with these generators, as shown in Fig. \ref{voro_schm} (c). For a cell in this Voronoi diagram, if an edge of the cell is infinite, it is marked with a dashed line; if it is finite, it is marked with a solid line.  In this diagram, we mainly focus on the red targeted particle and its corresponding Voronoi cell. If 
\begin{equation}
	\label{eq:voro3}	
	\left\| \boldsymbol{x}_{pi}  - \boldsymbol{x}_{ci}\right\| \ge \xi \Delta x
\end{equation}
is met, the red targeted particle will be detected as the surface particle. Here, it is found that  $ \xi=0.3 $ is suitable for this method. 

Therefore, if the red targeted particle is determined as the surface particle, it should satisfy at least one of the two conditions, i.e., the first condition is that one edge of the corresponding Voronoi cell of the targeted particle is infinite; 
the second condition is that the Voronoi cell is finite and Eq. (\ref{eq:voro3}) is met.
If both of the two conditions are not met, the particles will be determined as the interior particles. In practice, if the number of neighboring particles, $ N $, of the targeted particle is equal to the full number of neighboring particles, $ N_f $, the targeted particle will never be the surface particle. In order to reduce computational time, the determination process will be implemented only if the number of neighboring particles $ N $ is less than $ 0.9N_f $. 

The algorithm of the free-surface detection method is shown in Algorithm \ref{alg1}. In line 8, the `$ flag $' is used for recording whether the edge is infinite, and if the Voronoi cell edge is infinite, the loop of cell edges will break out and directly goes to line 25. In line 9, $S $, $S_x$ and $S_y$ are used to calculate the centroid $\boldsymbol{x}_c$ of the targeted Voronoi cell. 

In this study, the new free-surface detection method will be used, if not mentioned otherwise. The Boost polygon Voronoi library \cite{schaling2011boost} is used for Voronoi diagram implementations. \textcolor{black}{Constructing the 3D Voronoi diagram requires a well-established robust and efficient library, and thus the present framework is applied in 2D scenarios. The 3D realization will be our future work. } The performance of the new method is shown by challenging benchmark simulations in the validation section, i.e., the water impact on an elastic thin beam and the elastic structure impact on the free surface.

\begin{algorithm}
	\caption{Free-surface detection method based on Voronoi diagram }
	
	\begin{algorithmic}[1]
		
		\FOR{all particles}
		\STATE Identify neighboring particles within the radius of $ 1.8\Delta x $ and construct the array of generators  $Points(i)$;
		\STATE The targeted particle is set as a non-surface particle;
		
		\IF{Neighboring particle number $ N \le 0.9N_f $}
		\STATE Construct Voronoi diagram with generators  $Points(i)$;
		\FOR{Cells in the Voronoi diagram}
		\IF{Cell is corresponding to the targeted particle $ \boldsymbol{x}_p $}
		\STATE $flag \gets 0$;
		\STATE $S \gets 0$, $S_x \gets 0$, $S_y \gets 0$;
		\FOR{Edges in the corresponding cell}
		\IF{Edge is infinite}
		\STATE $flag \gets 1$;
		\STATE break;
		
		\ELSE
		\STATE $x_0, y_0 \gets \text{coordinates of first vertex of the edge}$;
		\STATE $x_1, y_1 \gets \text{coordinates of second vertex of the edge}$;		
		\STATE $S \gets S + 0.5(x_{0}y_{1}-x_{1}y_{0})$;
		\STATE $S_x \gets S_x + (x_{0}y_{1}-x_{1}y_{0})(x_{0}+x_{1})/6$;
		\STATE $S_y \gets S_y + (x_{0}y_{1}-x_{1}y_{0})(y_{0}+y_{1})/6$;
		
		\ENDIF
		
		\ENDFOR
		\IF{$ S \ne 0$ }
		\STATE Coordinates of cell centroid $\boldsymbol{x}_c$ $ \gets $ $S_x/S, S_y/S  $;
		\ENDIF
		
		\IF{$ \left\| \boldsymbol{x}_p  - \boldsymbol{x}_c\right\| \ge 0.3\Delta x $ or $flag=1$ }
		\STATE The particle is set as a surface particle;
		
		\ENDIF
		
		\ENDIF
		\ENDFOR
		
		\ENDIF
		
		\ENDFOR  
		
	\end{algorithmic}
	\label{alg1}
\end{algorithm}
\section{An improved APR method}
\label{sect3}

The present multi-resolution method, so-called APR proposed by Chiron et al. \cite{chiron2018analysis}, deploys a pre-defined refinement zone with a finer resolution while preserving the coarse resolution in the left areas. In
many scenarios, there are more than two resolution levels in the computational domains for heavily reducing the computational costs, as shown in Fig. \ref{APR_schm}(a), where the solid is resolved by the finest level. The set of particles in resolution level $k$ is denoted as $ L_k $.   In this method, the particles of different levels are either `active' or `inactive', as shown in Fig. \ref{APR_schm}(b): if the particle is `active', it will take part in the physical time evolution by solving the governing equations; if the particle is `inactive', the field values of the particle, such as density, velocity, and acceleration, are derived by the so-called Shepard interpolation method \cite{shepard1968two}. Considering two neighboring levels,  e.g., $ L_k $ and  $ L_{k+1} $: for $ L_{k} $ particles, if they enter $ L_{k+1} $ particle zone, they will be `inactive'; for $ L_{k+1} $ particles, if they exit from $ L_{k+1} $ particle zone, they will be `inactive', and if they further exit from the non-regularized transition sub-zone (discussed later), they will be deleted. 

In the present improved APR method, a large refinement scale ratio of 4 with multiple resolution levels will be devised to further decrease the total particle number and save computational costs. In order to eliminate the irregular particle distribution induced by the large refinement scale ratio in the transition zone, a regularized transition sub-zone is deployed to smooth the solution between two neighboring resolution levels. A multi-resolution data structure is also incorporated to enhance the efficiency of the improved APR method. 

\subsection{Large refinement scale ratio }
\begin{figure}[htbp]
	\centering
	\includegraphics[width=0.8\textwidth]{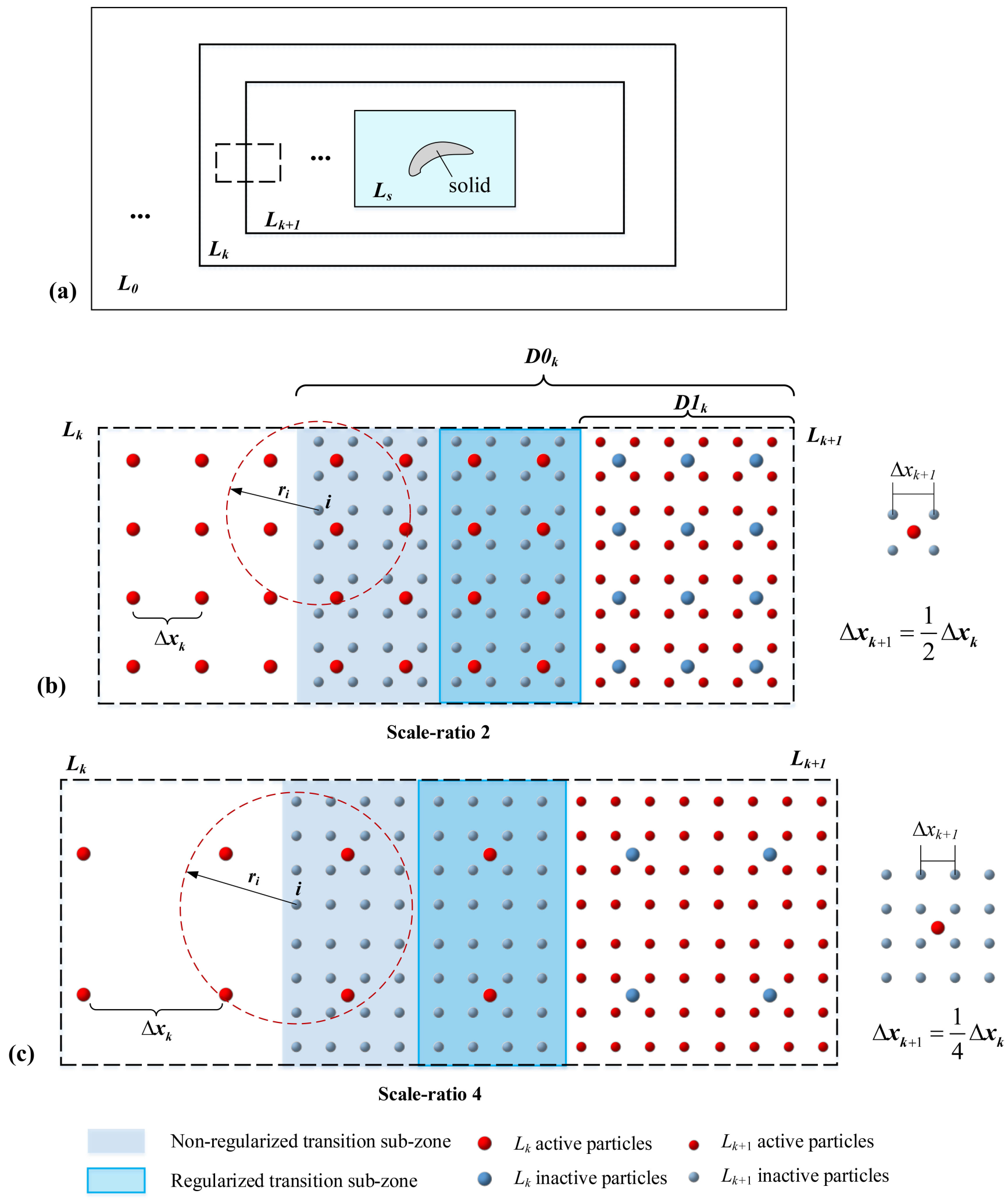}
	\caption{Schematic of the improved APR method: (a) the APR refinement zone is defined around the solid with multiple resolution levels; (b) and (c) display  the relationships for two neighboring refinement levels $ L_k $ and $ L_{k+1} $ for refinement scale ratio 2 and 4, where the interpolation radius $ r_i $ of the `inactive' particle is shown in red dashed lines, and the particle splitting graphs are shown on the right correspondingly. \textcolor{black}{Domain $D0_k$ and $D1_k$ represent the $L_{k+1}$ domain and the $L_{k+1}$ domain except the transition zone, respectively.} }
	\label{APR_schm}
\end{figure}

For the standard APR method proposed by Chiron et al. \cite{chiron2018analysis}, the scale ratio is 2, as shown in Fig. \ref{APR_schm}(b). If a $ L_k $ particle enters the pre-defined refinement zone, $4$ new $ L_{k+1} $ particles, with the $ L_k $ particle at the center, will be generated for 2D problems, and the relationships between the particle spacing $ \Delta x_{k+1} $, smooth length $ h_{k+1} $, mass $ m_{k+1} $ in level $ L_{k+1} $ and those parameters $ \Delta x_{k} $, $ h_{k} $,  $ m_{k} $ in level $ L_{k} $ are given as
\begin{equation}
	\label{eq:Lk2}	
	\begin{aligned}
		\Delta x_{k+1}=\frac{1}{2}\Delta x_{k}, \quad
		h_{k+1}=\frac{1}{2}h_{k}, \quad
		m_{k+1}=\frac{1}{4}m_{k}.
	\end{aligned}
\end{equation}
When the scale ratio is 2, many resolution levels can be used to refine the targeted areas of interest. In this way, the particles in the area of interest are very dense, and most of these particles are `inactive' particles, which are usually redundant in the refinement zone. On the other hand, if we choose the scale ratio as 4, the total number of particles in the computational domain can be greatly reduced, which will reduce the computational time sharply and save memory storage as well. The parameters of the newly generated particles using a scale ratio of 4 are
\begin{equation}
	\label{eq:Lk4}	
	\begin{aligned}
		\Delta x_{k+1}=\frac{1}{4}\Delta x_{k}, \quad
		h_{k+1}=\frac{1}{4}h_{k}, \quad
		m_{k+1}=\frac{1}{16}m_{k}.
	\end{aligned}
\end{equation}
\begin{algorithm}
	\caption{\textcolor{black}{Location determination of the `inactive' particle $i$ among all resolution levels}}
	
	\begin{algorithmic}[1]
		\STATE $l_0$=0,$\ $ $l_1$=0;
		\FOR{level $k$ from 0 to $N-1$ }
		\IF{$i$ $\in$ $D0_k$}
		\STATE $l_0$ ++;
		\ENDIF
		\IF{$i$ $\in$ $D1_k$}
		\STATE $l_1$ ++;
		\ENDIF		
		\ENDFOR  
		
	\end{algorithmic}
	\label{alg2}
\end{algorithm}

The sketch of the refinement with scale ratio 4 is shown in Fig. \ref{APR_schm}(c). \textcolor{black}{The interpolation radius $ r_i $ of the `inactive' particle $ i $ in Fig. \ref{APR_schm} is estimated based on the particle location among different levels. Two location indicators $l_0$ and $l_1$ are determined through looping in all the resolution levels. The corresponding algorithm is shown in Algorithm \ref{alg2}, where domain $D0_k$ and $D1_k$ are referred to Fig. \ref{APR_schm}, and the $N$ denotes the resolution levels.} Then the interpolation radius $ r_i $ is expressed as
\begin{equation}
	\begin{aligned}
		\textcolor{black}{r_i=\frac{1}{2}q(h_{l_0} + h_{l_1}) . }
	\end{aligned}
\end{equation}
\textcolor{black}{For example, if $L_0$ `inactive' particle is in $L_3$ refinement zone, the interpolation radius for the $L_0$ `inactive' particle is $q h_3$; if $L_0$ `inactive' particle is in $L_3$ transition zone, the interpolation radius for the $L_0$ `inactive' particle is $\frac{1}{2}q(h_2+h_3)$. This radius definition ensures that enough `active' particles can be found for the interpolation due to the employed large refinement scale ratio 4, and also ensures that the interpolation radius is not too large for a local interpolation.} This radius is different from $ r_i=qh_{k+1} $  in the standard APR. The field values of particle pressure, velocity, and acceleration are derived with the Shepard interpolation function $ f \left( \phi \right) $ with regard to the field value $ \phi $, which is given as
\begin{equation}
	\label{eq:shepard}
	\begin{aligned}
		&f \left( \phi \right) = \sum_{j \in \Omega_{i}} \phi_j W_{i j} V_{j}/\sum_{j \in \Omega_{i}} W_{i j} V_{j}, 
	\end{aligned}
\end{equation}
where `$ j $' is the `active' particle within the support domain $ \Omega_{i} $ of radius $ r_i $.

\subsection{Refinement transition zone with a regularized transition sub-zone for large refinement scale ratio}
Apart from the large refinement scale ratio, a regularized transition sub-zone between two neighboring levels is supplemented for better numerical robustness. The transition zone in the APR method works as an information exchange channel between two neighboring levels and provides full kernel support for all the `active' particles. Usually, when the $ L_k $ particles enter the transition zone, the distribution of new $ L_{k+1} $ particles generated by the $ L_k $ particles is non-isotropic, especially for the large refinement scale ratio, where the newly generated particles are very likely to cluster, and this will degenerate the accuracy of the following SPH simulations. Indeed, the field values of `inactive' particles are derived with  the Shepard interpolation, therefore the isotropic particle distribution is the only requirement for those particles. 
In order to render an isotropic particle distribution of the  `inactive' particles, we use a modified background driving force $ \boldsymbol{\nabla} Pb_{i\in L_m} $ to regularize these particles \textcolor{black}{\cite{monaghan2000sph}\cite{sun2017deltaplus}}. For $ L_m $ `inactive' particle $ i $, the background driving force $ \boldsymbol{\nabla} Pb_{i\in L_m} $ is given as
\begin{equation}
	\label{eq:background_P1}
	\begin{aligned}
		&\boldsymbol{\nabla} Pb_{i\in L_m} = -\rho_{0} c_{0}^2 \sum_{j \in L_m} \left[1+0.2\left(\dfrac{W_{i j}}{W(\Delta x_m,h_m)}\right)^{4}  \right] \boldsymbol{\nabla}_{i} W_{i j} V_{j}, 
	\end{aligned}
\end{equation}
where the neighbor particle $ j \in L_m $ comes from the same level $L_m $. This formula ensures a sufficient driving force to regularize the `inactive' particles for a large scale ratio. 

For those particles at the outer layer of the transition zone, their kernel support is incomplete, and Eq. (\ref{eq:background_P1}) cannot be directly deployed. Hence, the  transition zone is divided into the regularized transition sub-zone and the non-regularized transition sub-zone, as shown in Fig. \ref{APR_schm}. In the  regularized transition sub-zone, adjacent to the  $ L_{k+1} $ `active' particles, the particle distribution will be regularized with Eq. (\ref{eq:background_P1}); in the  non-regularized transition sub-zone, the particle distribution will not be regularized. The thickness of the non-regularized transition sub-zone is chosen as $ 4\Delta x_{k+1} $ to involve all the `inactive' particles of which the kernel support is incomplete. A thickness of   $ 4\Delta x_{k+1} $ is chosen for the regularized transition sub-zone.
In fact, all the `inactive' particles, except for the particles in the non-regularized transition sub-zone and surface area, will be regularized with Eq. (\ref{eq:background_P1}). The performance of deploying the regularized transition sub-zone will be shown in section \ref{subsect4.1}.

For the $ L_{m} $ `inactive' particles, the evolution of field values are calculated in sequence as
\begin{equation}
	\label{eq:regular_inact}
	\begin{array}{l}
		\begin{aligned}
			&\boldsymbol{v}^{n+\frac{1}{2}} = \boldsymbol{v}^{n} + \frac{\Delta t_{F}}{2} f \left( \frac{d \boldsymbol{v}}{dt} \right) ,\\
			&\boldsymbol{\tilde{v}}^{n+\frac{1}{2}} = \boldsymbol{v}^{n+\frac{1}{2}} + \frac{\Delta t_{F}}{2\rho_0}\boldsymbol{\nabla} Pb^n_{i\in L_m},\\
			&\boldsymbol{x}^{n+1} = \boldsymbol{x}^{n} + {\Delta t_{F}} \boldsymbol{\tilde{v}}^{n+\frac{1}{2}},\\
			&\rho^{n+1} = f \left( \rho \right),\\
			&\boldsymbol{v}^{n+1} =f\left( \boldsymbol{v} \right).\\
		\end{aligned}
	\end{array}
\end{equation}
In this study, Eq. (\ref{eq:scheme1}) is adopted for the evolution of the `active' particles, while Eq. (\ref{eq:regular_inact}) is used for the `inactive' particles. 

\textcolor{black}{In the work of Chiron et al. \cite{chiron2018analysis}, the particle disordering method is used, however, there are some differences between the present improved version and that of Chiron et al. \cite{chiron2018analysis}. Firstly, the regularized transition sub-zone is deployed between the coarse and fine levels to ensure a much smoother pressure distribution, and we also use a larger transition zone thickness to ensure better smoothness; secondly, a much stronger background driving force, i.e., Eq. (\ref{eq:background_P1}), is used for the `inactive' particles to ensure a more isotropic distribution; thirdly, the particle regularization for the refinement scale ratio 2 and 4 is designed in the same APR framework; furthermore, in the work of Chiron et al. \cite{chiron2018analysis}, the particle regularization implementations between the coarse and fine levels are not elaborated clearly, and as such we give much more detailed descriptions to tackle the particle regularization between the interfaces of different levels. }

\subsection{Multi-resolution data structure}
\begin{figure}[htbp]
	\centering
	\includegraphics[width=0.8\textwidth]{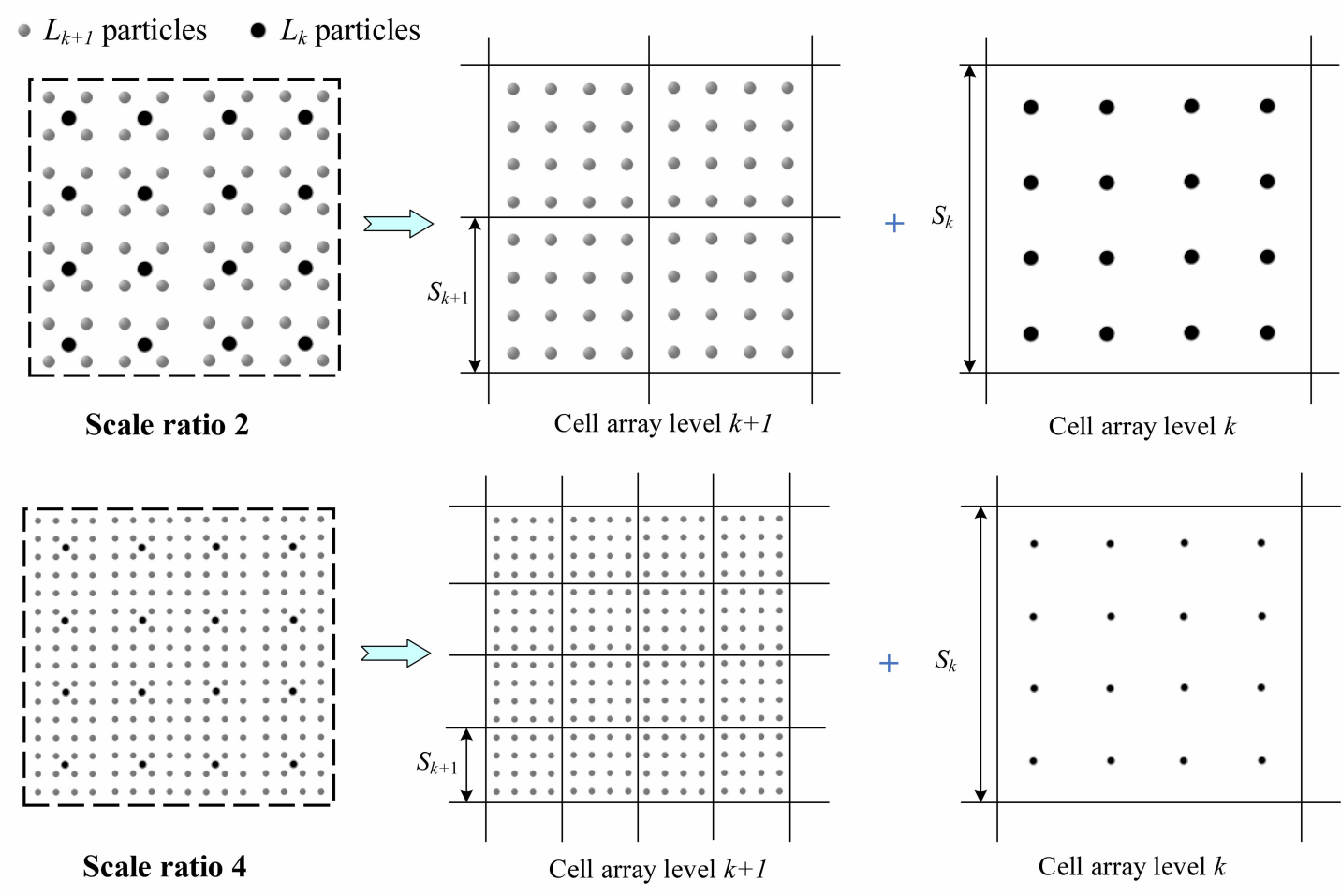}
	\caption{Multi-resolution data structure for improved APR: the `gray' and `black' particles represent $  L_{k+1} $ and $ L_k $ particles. The particles from different levels are mapped to the corresponding cell of different levels, and the cell scale   $ S_n $ at level $ n $ is set as $ S_n=4\Delta x_n $. }
	\label{APR_schm_cell}
\end{figure}

In the above APR method, there are more than one resolution level, and the data structure for the particles at different levels is important for efficient neighbor searching. Fu et al. \cite{fu2019parallel} and Ji et al. \cite{ji2019new} proposed a multi-resolution data structure for SPH particles based on the multi-resolution cell-linked list method. In this method, the particles from different levels are mapped to the corresponding cell of different levels, as shown in Fig. \ref{APR_schm_cell}. In level $ k $, the scale of cell $ S_k $ is defined as $ S_k=4\Delta x_k $ considering the chosen particle smoothing length in this study, and $ S_k/S_{k+1}$ equals 2 for the refinement scale ratio 2 and $S_k/S_{k+1}$ equals 4 for the scale ratio 4. In this way, the neighboring particles within the kernel support can be searched with high efficiency, even in a distributed-memory parallel environment. For more details, the readers are referred to \cite{fu2019parallel}\cite{ji2019new}.

\textcolor{black}{In all, it is worth noting that the refinement scale ratio is limited within 4 in this work. A larger scale ratio, e.g., 8, may degenerate the solution accuracy due to the very large interpolation radius in the transition zone, and therefore we do not strive to extend it to refinement scale ratio 8. }

\section{Numerical validations}
\label{sect4}
In this section, a set of benchmark cases will be simulated with the improved APR method, and the performance of the novel free-surface detection method based on the Voronoi diagram will also be assessed thoroughly. The cases considered include water impact on an elastic thin beam, water impact on an elastic thin beam with the two-phase problem, elastic structure impact on the free surface and fishlike swimming. 

\subsection{Water impact on an elastic thin beam}
\label{subsect4.1}
\begin{figure}[htbp]
	\centering
	\includegraphics[width=0.6\textwidth]{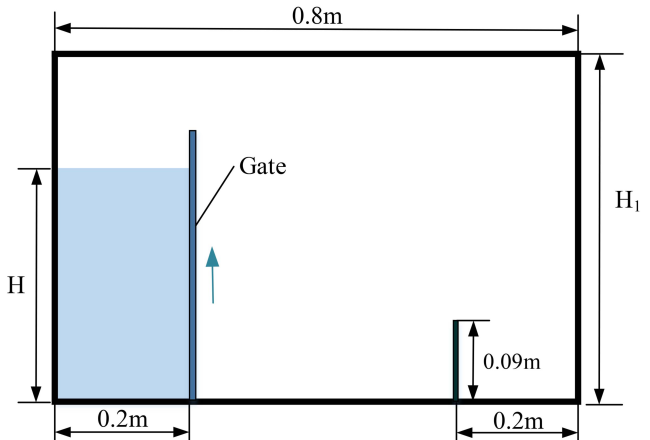}
	\caption{Schematic of water impact on an elastic thin beam.}
	\label{schm_beam}
\end{figure}

This case is initially set up by Liao et al. \cite{liao2015free}. The setup parameters and dimensions of this case are shown in Fig. \ref{schm_beam}. The water height is $ H= 0.4 \text{ }\rm m $, and the computational domain height $ H_1 $ should be larger than $ H $. When the gate opens by going upward, the water will flow outside and impact onto the rubber beam on the right. The water density is $ 997 \text{ }\rm {kg/m^3} $ and is assumed to be inviscid in this case. The gravity acceleration is $ g=9.81 \text{ }\rm{m/s^2} $. The artificial sound speed is chosen as $c_0= 20\text{ }\rm {m/s} $. The rubber beam is very thin with $ T=0.004\text{ }\rm m $ in thickness. The density, Young's modulus and Poisson's ratio of the rubber beam are $ \rho_{s}=1161.54 \text{ }\rm {kg/m^3} $, $ E_s=3.5 \text{ }\rm {MPa} $ and $ \nu=0.45 $. The motion of the gate on the right of the water column is referred to \cite{sun2019study}. The Mooney-Rivlin constitutive model is employed for the rubber beam, and the three parameters \cite{marino2021numerical} in Eq. (\ref{mooney}) are $ C_{10}=5.98\times 10^5 \text{ \rm{Pa}} $, $ C_{01}=0 \text{ \rm{Pa}} $ and $ D_1=8.57\times 10^{-8} $. The viscosity term Eq. (\ref{visco}) is used here for the solid damping, and kinematic viscosity $\eta/\rho$ is chosen as 0.025. It is worth noting that this is an artificial term for the damping effect.
In most cases, at least four particles \textcolor{black}{(will be more if acting as fluid boundary)} should be deployed to resolve the thickness of the beam, and if a uniform particle resolution is chosen, the total number of particles arranged in the entire domain will be extremely large, resulting in the increase of unnecessary computational costs. For this problem, an APR refinement zone is deployed around the beam. Specifically, the finest resolution is  $ \Delta x^s=T/6$, and simulations of (i) the refinement scale ratio 4 with 2 levels and (ii) the refinement scale ratio 2 with 3 levels are conducted and compared.

\begin{figure}[htbp]
	\centering
	\includegraphics[width=0.9\textwidth]{ 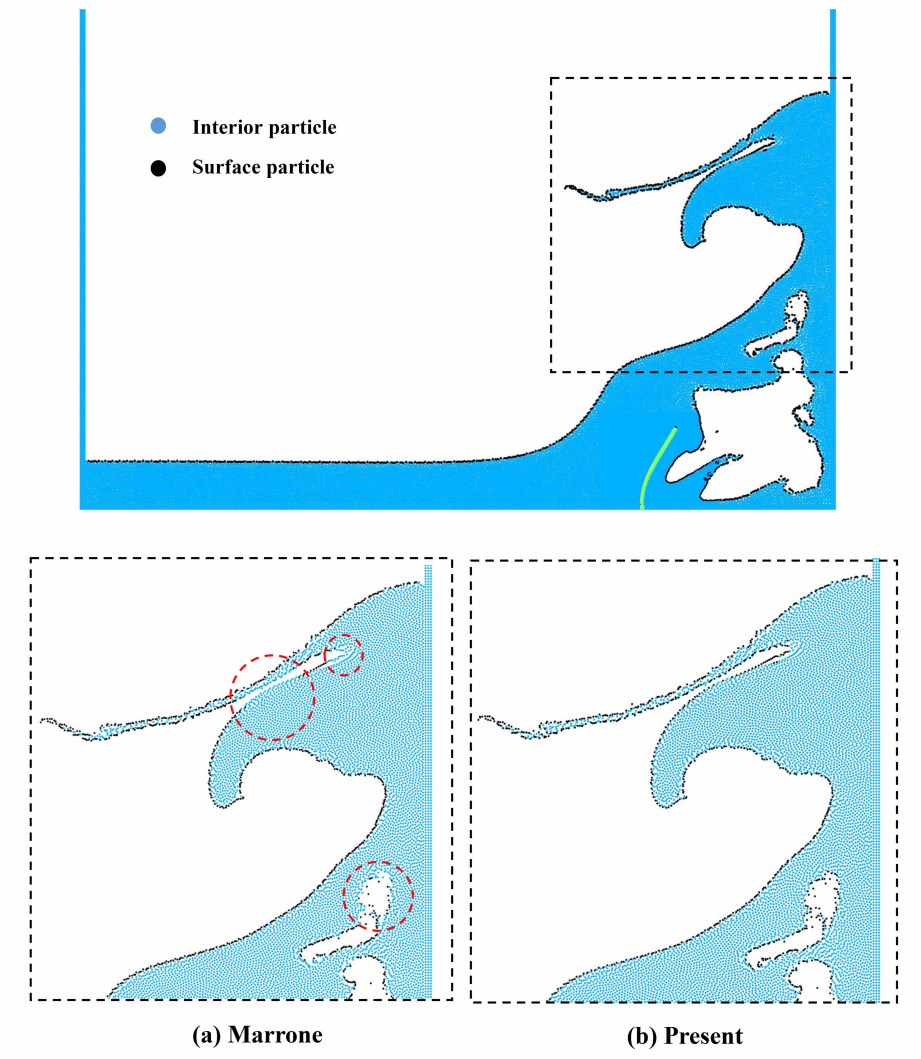}
	\caption{The surface particles are detected with the conventional method \cite{marrone2010fast} and the present method, and the zoomed-in details in (a) and (b) display the comparison between the two methods. The circle areas (red dashed line) indicates the drawbacks of the conventional method. }
	\label{voro_surf}
\end{figure}

The surface particles (marked with blue color) detected by the proposed method are shown in Fig. \ref{voro_surf}. This snapshot is captured from the latter stage of impact flow, which reveals a rather complex surface topology. The conventional method based on solving the eigenvalue problem proposed by Marrone et al. \cite{marrone2010fast} is compared with the present method based on the Voronoi diagram. In general, the present method is capable of detecting all surface particles satisfactorily, which is consistent with the conventional method. Moreover, the present method can even detect some surface particles that are missed by the conventional method. For example, when two fluid surfaces are very close to each other (circles in dashed line in Fig. \ref{voro_surf}(a)), the particles in this area cannot be detected by the conventional method, because the particles from the opposite surface area will be in the so-called scan region of the concerned particles, resulting in an incorrect detection. In contrast, the present method can successfully detect these particles quite well, as shown in Fig. \ref{voro_surf}(b). The computational cost of the present method is shown in Table \ref{tb:voro_surf}. The overall computational time of each time step evolution is $ 0.109 \text{ } \rm s $. It shows that the computational cost of the present method for the free-surface detection only accounts for $ 1.5\% $ of the total computational time for each time step evolution,  which is negligible.

\begin{table}
	\renewcommand{\arraystretch}{1.3}
	\caption{Comparison of  computational costs of the free-surface  detection methods between the conventional method and the present method.} 
	\begin{center}
		%%% Some packages, such as MDW tools, offer better commands for making tables
		%%% than the plain LaTeX2e tabular which is used here.
		\begin{tabular}{p{6cm} p{4cm} p{3cm}}
			\toprule [1.2 pt]
			& Marrone et al. \cite{marrone2010fast} & Present \\
			\hline
			Computational time (s) & 0.003 & 0.0017 \\			
			Proportion in each time step & $ 2.7\% $ &  $ 1.5\% $  \\			
			\bottomrule [1.2 pt]
		\end{tabular}
	\end{center}
	\label{tb:voro_surf}
\end{table}

\begin{figure}[htbp]
	\centering
	\includegraphics[width=0.8\textwidth]{ 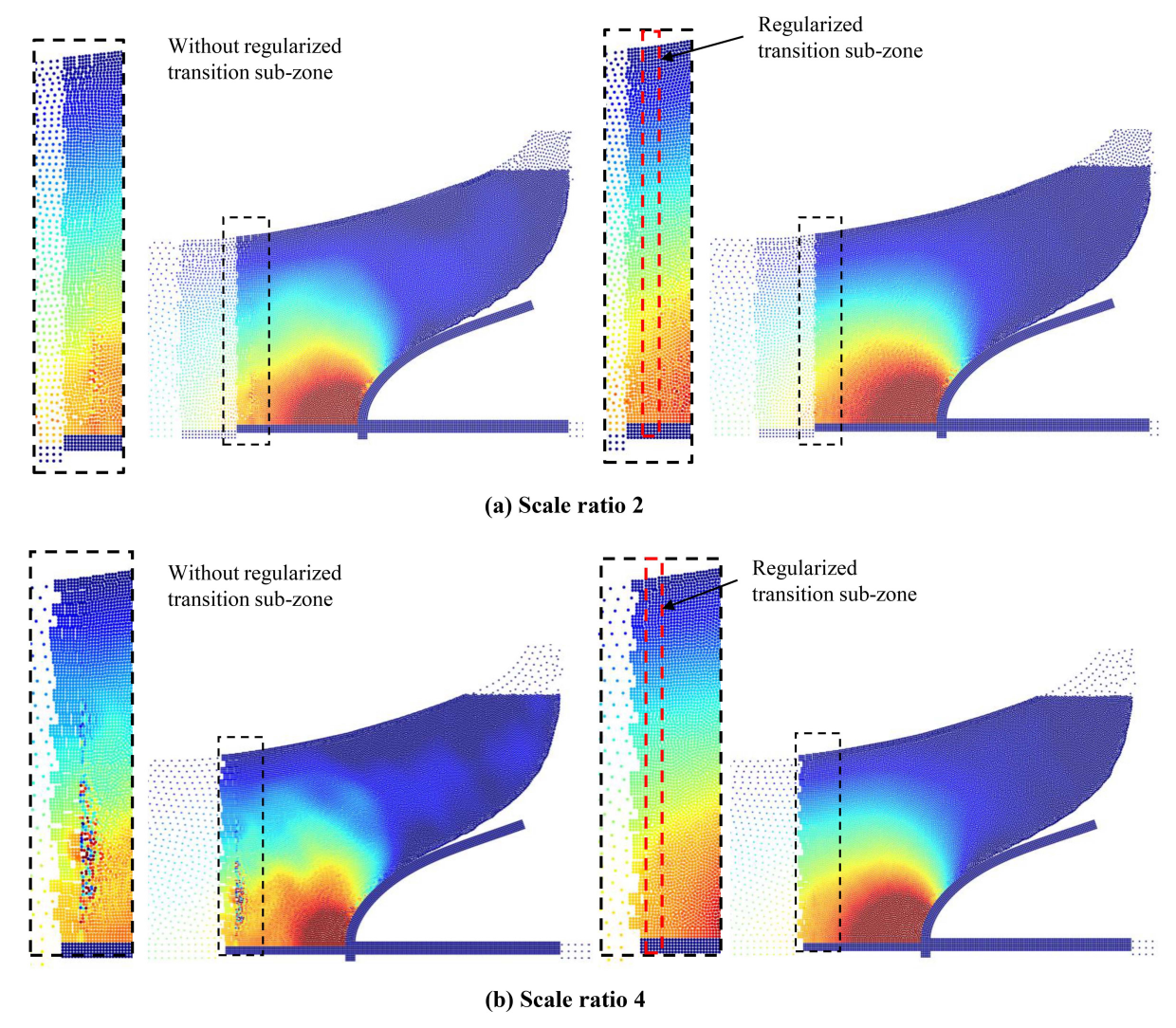}
	\caption{\textcolor{black}{The regularized transition sub-zone and the pressure fields: the regularized transition sub-zone is used for the simulation with the refinement scale ratio (a) 2 and (b) 4. The left shows that no regularized transition sub-zone is deployed for the case with the refinement scale ratio 4, and the pressure field is extremely unstable near the transition zone; the right shows that the regularized transition sub-zone (in red dashed line) is deployed, and the pressure distribution is smooth across different resolutions.} }
	\label{regu_tran_subzone}
\end{figure}

\textcolor{black}{ Fig. \ref{regu_tran_subzone} shows the pressure fields solved without and with the regularized transition sub-zone using the refinement scale ratios 2 and 4. When the regularized transition sub-zone is not deployed, as shown in the left column of Fig. \ref{regu_tran_subzone}, the pressure distribution across different resolutions is extremely noisy near the transition zone, especially for the case with scale ratio 4; when the regularized transition sub-zone is deployed, as shown in the right column of Fig. \ref{regu_tran_subzone}, the particle distribution is isotropic in this area, and the pressure distribution is very smooth across different resolutions. Therefore, the deployed regularized transition sub-zone is very useful to increase the numerical robustness between different resolutions, especially for a large refinement scale ratio of 4.}

\begin{figure}[htbp]
	\centering
	\includegraphics[width=1\textwidth]{ 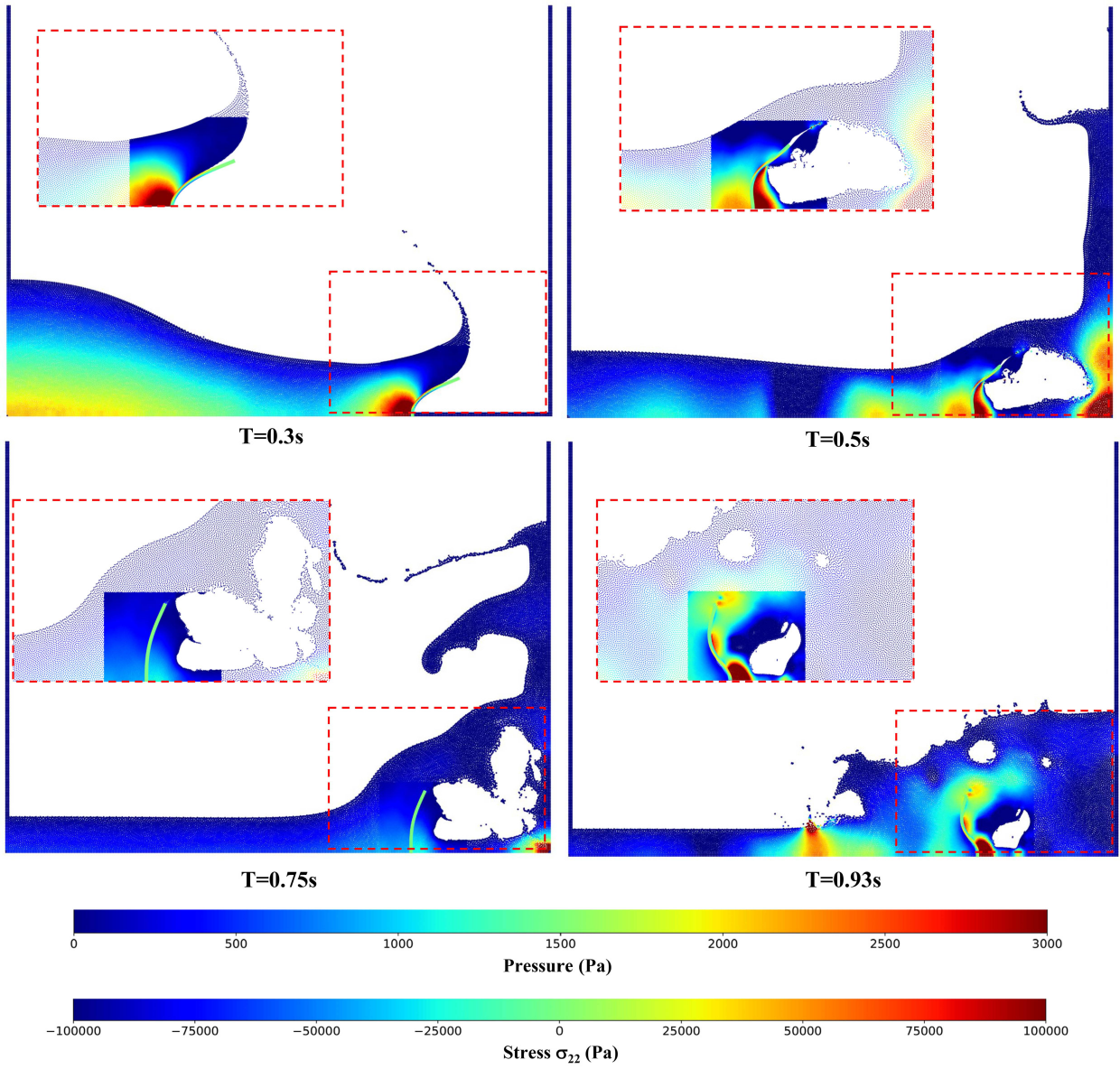}
	\caption{Water impact on an elastic thin beam: the pressure  and Cauchy stress component $\sigma_{22}$ in vertical direction are shown at different time, with zoomed-in details displayed in the red dashed lines.}
	\label{beam_snap}
\end{figure}

\begin{figure}[htbp]
	\centering
	\includegraphics[width=0.9\textwidth]{ 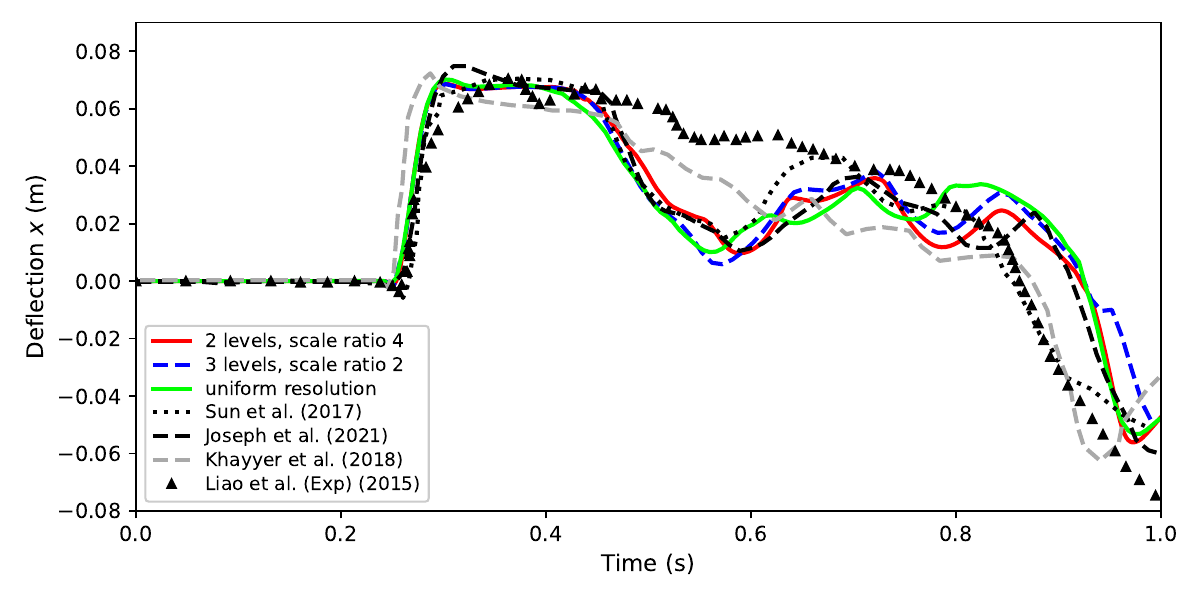}
	\caption{Predicted deflection of the rubber beam  at $ 0.0875 \text{ }\rm m $ from simulations (i) and (ii)  and its comparisons with the results from single-phase SPH by Sun et al. \cite{sun2019study}, single-phase MPS by Khayyer et al. \cite{khayyer2019multi}, single-phase 3D SPH by Joseph et al. \cite{o2021fluid} and experimental data by Liao et al. \cite{liao2015free}.}
	\label{defl_plt_1p}
\end{figure}

The evolution of the water impact using the refinement scale ratio 4 with 2 levels is illustrated in Fig. \ref{beam_snap}, where the pressure field of water and the stress component in the vertical direction are shown. Thanks to the regularized transition sub-zone deployed, the pressure field has a very smooth transition from the non-refinement zone to the refinement zone. Fig. \ref{defl_plt_1p} displays the predicted deflection of the rubber beam from simulations (i) and (ii) at $ 0.0875 \text{ }\rm m $ and its comparisons with the results from single-phase SPH by Sun et al. \cite{sun2019study}, single-phase MPS by Khayyer et al. \cite{khayyer2019multi}, single-phase 3D SPH by O’Connor et al. \cite{o2021fluid} and the experiment by Liao et al. \cite{liao2015free}. For simulations (i) and (ii), the two results almost overlap before $ 0.45 s $; after $ 0.45 s $, though there is a marginal deviation, the agreement is still quite good. Comparably, the present results are quite consistent with those by Sun et al. \cite{sun2019study} and Joseph et al. \cite{o2021fluid}. It is observed that almost all the results deviate from the experiments at around $ 0.6s $, which is attributed to the neglected air trapping effect due to the single-phase model employed in the simulation. This little divergence is under expectation because the flow is very violent in the impact case and the simulation result is very sensitive to the particle distribution.

Table \ref{tb:beam} shows the average computational time for each 0.002s between simulations (i) and (ii). The simulation (ii) using 2 levels with the refinement scale ratio of 4 saves nearly $ 30\% $ computational cost when compared to the simulation (ii) using 3 levels with the refinement scale ratio of 2. Moreover, the computational cost with the refinement scale ratio of 4 is only about 1/8 of that with the uniform fine resolution. This comparison demonstrates that the improved APR method with a large refinement scale ratio can preserve the simulation accuracy and reduce computational costs substantially. 
\begin{table}
	\renewcommand{\arraystretch}{1.3}
	\caption{Comparison of the average computational time for each 0.002 s between simulations (i), (ii) and using uniform resolution.} 
	\begin{center}
		%%% Some packages, such as MDW tools, offer better commands for making tables
		%%% than the plain LaTeX2e tabular which is used here.
		\begin{tabular}{p{2cm} p{5cm} p{5cm} p{2cm}}
			\toprule [1.2 pt]
			Resolution & (i): 3 levels,  scale ratio 2 
			& (ii): 2 levels, scale ratio 4 & Uniform \\
			\hline
			$ \Delta x^s=T/6$ & 34\text{ }s & 24\text{ }s  & 188 s\\						
			\bottomrule [1.2 pt]
		\end{tabular}
	\end{center}
	\label{tb:beam}
\end{table}
\subsection{Water impact on an elastic thin beam with the two-phase model}

In the case above, only the water phase is considered based on a single-phase model. When the water jet impacts the right wall, there is a trapped area. Owing to the deployed single-phase model, the trapped area is treated as a vacuum, which is not consistent with real physics. The trapped air bubbles usually have a non-negligible effect on the flow evolution. Therefore, we will investigate this challenging FSI problem with a two-phase flow model. Using the same setup in Fig. \ref{schm_beam}, the air phase is deployed in the area apart from water, and the density of air is set as $ 1.29 \text{ }\rm {kg/m^3} $. The background pressure $ p_{bg} $ is employed in the two-phase model for stability, and it is chosen as $ p_{bg}=1000 \text{ }\rm{Pa} $. The artificial sound speed is chosen as $c_0= 50\text{ }\rm {m/s} $. The artificial viscosity coefficient is set as $\alpha=0.1 $ \cite{sun2019study}.  In this case, the APR method with 2 refinement levels and the scale ratio 4 is used firstly, and \textcolor{black}{ the finest resolution is set as $ \Delta x^s=T/6$. }

\begin{figure}[htbp]
	\centering
	\includegraphics[width=1\textwidth]{ 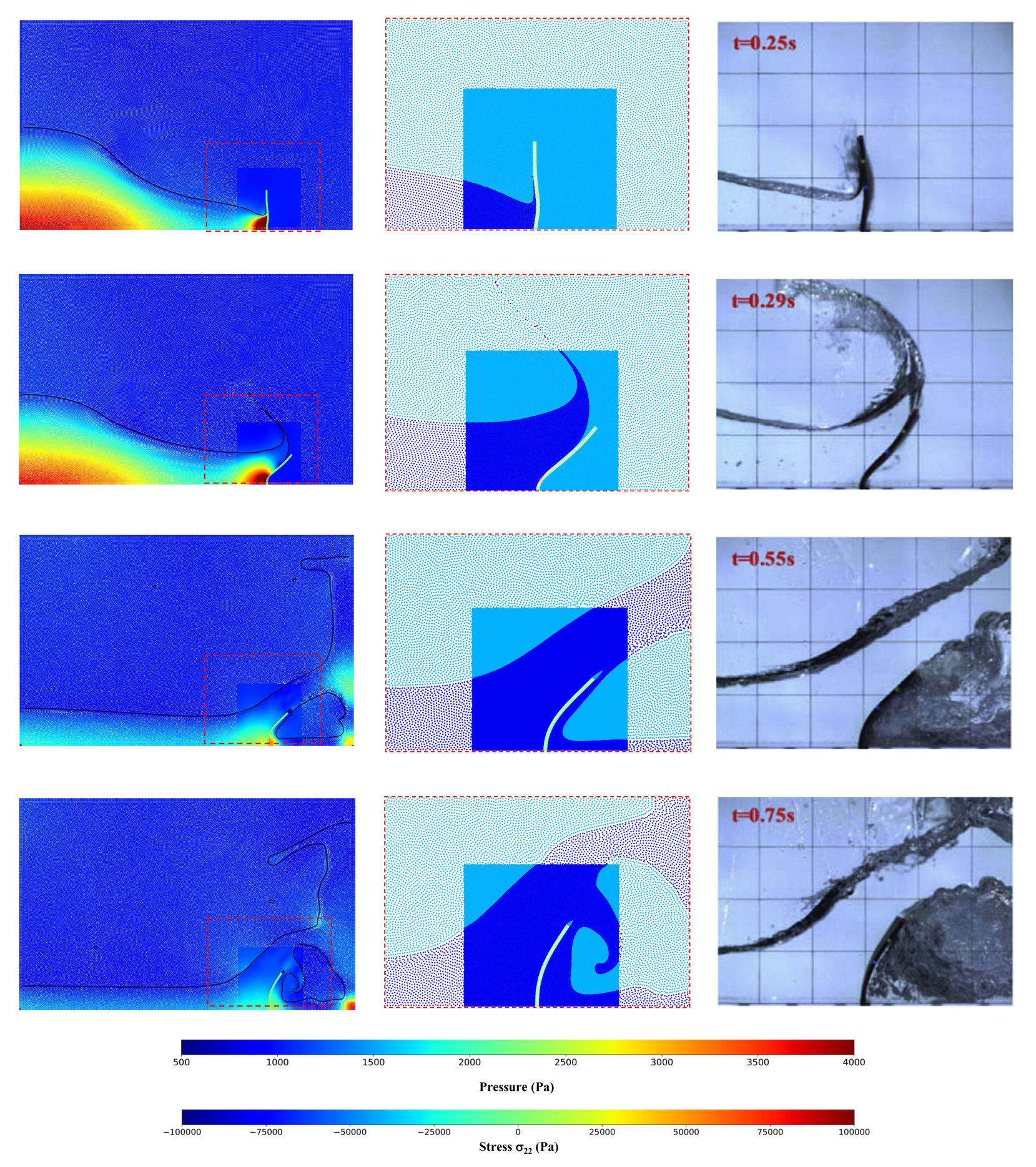}
	\caption{\textcolor{black}{Water impact on an elastic thin beam with the two-phase simulation: the distributions of pressure and Cauchy stress component $\sigma_{22}$ in the vertical direction at different simulation time are shown on the left. The zoomed-in details in the middle display the phase fields of the regions enclosed by the red dashed lines. The results are compared to the experimental snapshots by Liao et al. \cite{liao2015free} on the right.}}
	\label{multip_beam_snap}
\end{figure}

\begin{figure}[htbp]
	\centering
	\includegraphics[width=1\textwidth]{ 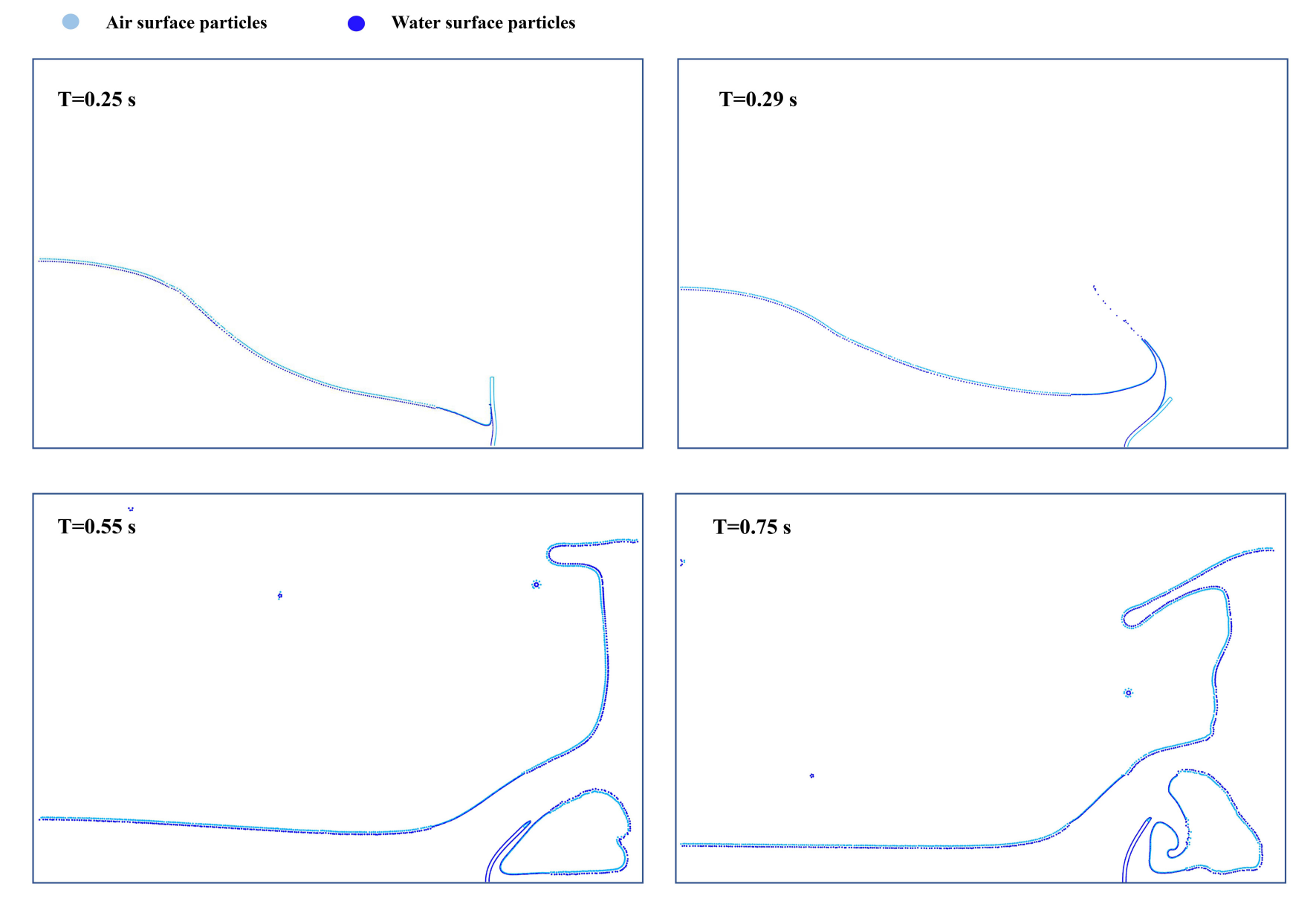}
	\caption{\textcolor{black}{Surface particle distributions of the two-phase problem, where the surface particles are identified by the new free-surface  detection method based on Voronoi diagram.}}
	\label{multip_surf}
\end{figure}

The snapshots of the simulation at different physical time are displayed in Fig. \ref{multip_beam_snap}. On the left, the pressure field and the Cauchy stress component in the vertical direction at different time are shown, where the interfaces of the two phases are depicted with black lines. It is observed that the pressure field is smooth across the phase interfaces and between the refinement zone and the non-refinement zone. In the middle, the zoomed-in details show that the phase interfaces are sharp without particle penetration. The simulation is compared to the experimental snapshots by Liao et al. \cite{liao2015free} shown on the right, and one can observe that the present results are very consistent with the experiments at different time. Furthermore, the surface particles are identified using the new free-surface  detection method, as shown in Fig. \ref{multip_surf}, and the interfaces of the two phases are detected successfully throughout the simulation. 

\begin{figure}[htbp]
	\centering
	\includegraphics[width=0.9\textwidth]{ 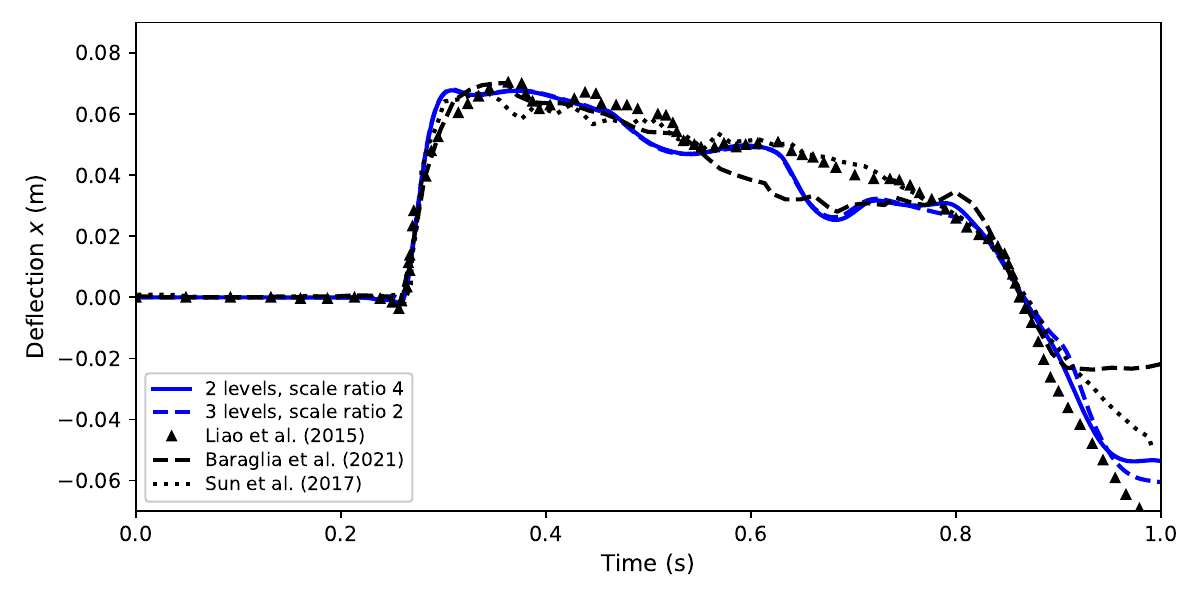}
	\caption{\textcolor{black}{Predicted deflection of the rubber beam  at $ 0.0875 \text{ }\rm m $ from simulations (i) and (ii), and its comparisons with the results from the two-phase corotational FEM by Baraglia et al. \cite{baraglia2021corotational}, the two-phase SPH by Sun et al. \cite{sun2019study} and the experimental data by Liao et al. \cite{liao2015free}.} }
	\label{multiP_0.4}
\end{figure}

The deflection of the rubber beam at the height of $ 0.0875 \text{ }\rm m $ is displayed in Fig. \ref{multiP_0.4}, and results of two simulation setups, i.e., 2 refinement levels with the scale ratio 4, and 3 refinement levels with the scale ratio 2, are compared. Firstly, it is shown that the two simulation results have quite good agreement throughout the simulation, indicating that the APR method with a large refinement scale ratio is reliable. Secondly, the present results are consistent with the experimental data by Liao et al. \cite{liao2015free}, despite a little deviation at around $ 0.65 \text{ } \rm s $, which is also observed in the results simulated with the two-phase corotational FEM by Baraglia et al. \cite{baraglia2021corotational}. In the latter stage, the present results agree better with the experimental data than that by Sun et al. \cite{sun2019study}. The abrupt decrease of deflection in the single-phase simulation is not observed in the two-phase simulations, owing to the deployed two-phase model considering  the effect of trapped air bubbles, as shown in Fig. \ref{multiP_0.4}. 

\begin{figure}[htbp]
	\centering
	\includegraphics[width=0.9\textwidth]{ 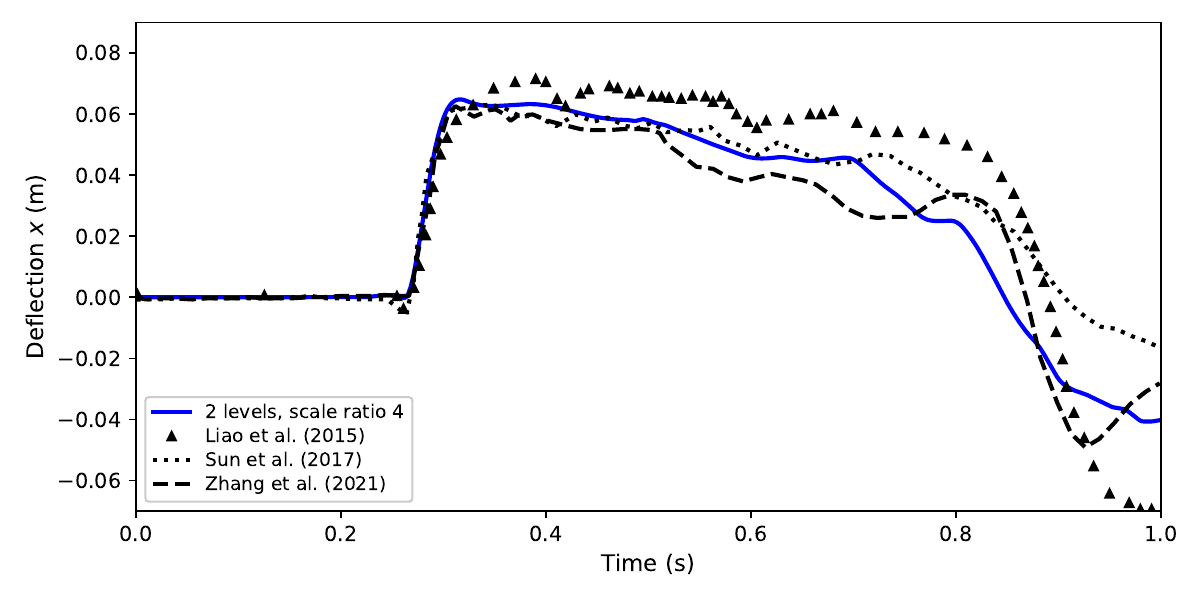}
	\caption{\textcolor{black}{Predicted deflection of the rubber beam at $ 0.0875 \text{ }\rm m $ from the simulation by 2 refinement levels with scale ratio 4 for $ H=0.3 \text{ } \rm m $, and its comparisons with the results from the two-phase SPH by Sun et al. \cite{sun2019study}, the two-phase Riemann-based SPH by Zhang et al. \cite{zhang2022efficient} and the experimental data by Liao et al. \cite{liao2015free}.}}
	\label{multiP_0.3_plt}
\end{figure}

Another case with the water height of $  H=0.3  \text{ }\rm m $ is further simulated using 2 refinement levels and the scale ratio 4, and the deflection of the beam is plotted in Fig. \ref{multiP_0.3_plt}. The results are compared with those from the two-phase SPH simulation by Sun et al. \cite{sun2019study}, the two-phase Riemann-based SPH simulation by Zhang et al. \cite{zhang2022efficient} and the experiment by Liao et al. \cite{liao2015free}. All these results are slightly lower than the experimental data which could be owing to the limitations of the 2D situation, and the present results are quite close to these studies. Table \ref{tb:beam1} shows the comparison of computational costs between simulation (i), (ii) and that with uniform resolution for each 0.002 s. It is observed that the computational time with the refinement scale ratio 4 reduces almost 25\%   when compared to that with the refinement scale ratio 2, and is approximately 1/8 of that using uniform resolution in this case. This case further validates the high performance of the present improved APR method with a large refinement scale ratio and the new free-surface detection method.

\begin{table}
	\renewcommand{\arraystretch}{1.3}
	\caption{\textcolor{black}{Comparison of the average computational time for each 0.002s between simulations (i), (ii) and that using uniform resolution, for $H=0.4$ m.} }
	\begin{center}
		\begin{tabular}{p{2cm} p{5cm} p{5cm}p{2cm}}
			\toprule [1.2 pt]
			Resolution & (i): 3 levels, scale ratio 2 & (ii): 2 levels, scale ratio 4 & Uniform\\
			\hline
			$\Delta x^s=T/6$ & 170 s & 135 s & 1017 s \\
			\bottomrule [1.2 pt]
		\end{tabular}
	\end{center}
	\label{tb:beam1}
\end{table}
\subsection{Elastic structure impact on the free surface}
\label{sect:imp}
\begin{figure}[htbp]
	\centering
	\includegraphics[width=0.6\textwidth]{ 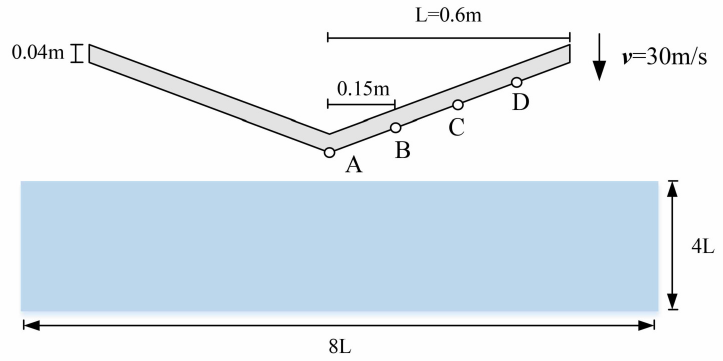}
	\caption{Schematic of an elastic structure impact on the free surface.}
	\label{schm_water_entry}
\end{figure}

The schematic of an elastic structure impact on the free surface is shown in Fig. \ref{schm_water_entry}. In this case, a wedge-shape elastic aluminum structure impacts on the water surface vertically with a constant velocity of $ v=30  \text{ }\rm{m/s} $. In the computational setup, the vertex $ A $ and two edges of the wedge are set at constant velocity of $ v=30  \text{ }\rm{m/s} $, and the two edges have no rotation. The wedge's half horizontal span is  $ L=0.6  \text{ }\rm{m} $ and its thickness is $T=0.04 \text{ } \rm{m} $. The water domain is $ 8L $ in width and $ 4L $ in height. The material properties of the aluminum wedge are defined as, i.e., Young's modulus $ E=67.5  \text{ }\rm{GPa} $, density $ \rho_s = 2700  \text{ }\rm{kg/m^3} $ and Poisson's ratio $ \nu=0.34 $. The sound speed is approximately $ 5000 \text{ }\rm{m/s}. $ Regarding the water phase, the density is $ \rho_w = 1000  \text{ }\rm{kg/m^3} $, and its sound speed is approximately $ 1500 \text{ }\rm{m/s}$. A larger artificial viscosity coefficient $ \alpha=0.4 $ and the density reinitialization technique \cite{colagrossi2003numerical} are used to smooth the pressure field and stabilize the water jet.  The gravity acceleration is $ g=9.81 \text{ } \rm{m/s^2} $. In this case, the artificial sound speed $ c_0 $ for fluid is set as the real sound speed of water.  Four probe locations, A, B, C and D, are arranged with equal distance along the half span of the wedge to probe the pressure history. Two simulations with different refinement scale ratios and resolution levels are implemented for this case. As for simulation (i), 4 refinement resolution levels are deployed and the scale ratio is set as 2; for simulation (ii), 3 refinement resolution levels are deployed and the scale ratio is set as 4. In both simulations, the resolution for the structure is set as $\Delta x^s =T/8 $. 

\begin{figure}[htbp]
	\centering
	\includegraphics[width=1\textwidth]{ 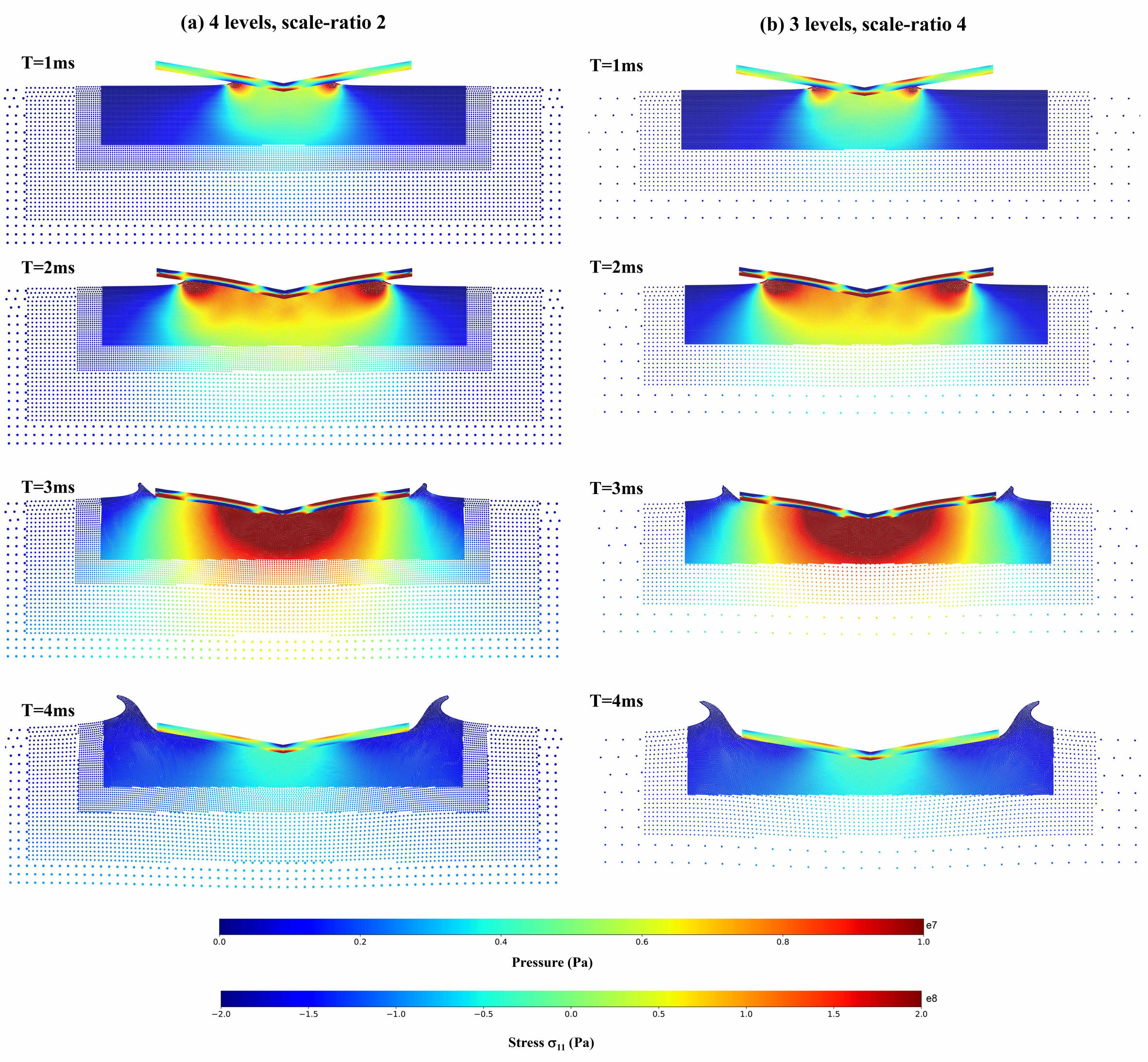}
	\caption{Elastic structure impact on the free surface: the pressure and Cauchy stress tensor $ \sigma_{11} $ are presented. The left column is from simulation (i), deploying 4 resolution levels and scale ratio 2; the right column is from simulation (ii), deploying 3 resolution levels and scale ratio 4. }
	\label{wedge_snap}
\end{figure}

\begin{figure*}[htbp]
	\centering
	
	\subfigure[Wedge deflection at location C.]{
		\includegraphics[width=0.7\linewidth]{ 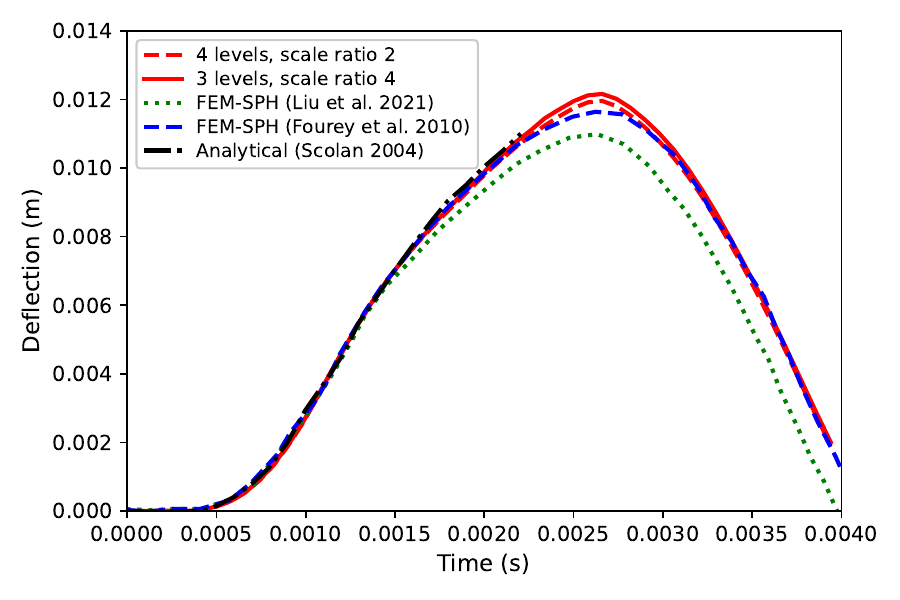}}
	\subfigure[Vertical force applied on the wedge.]{
		\includegraphics[width=0.7\linewidth]{ 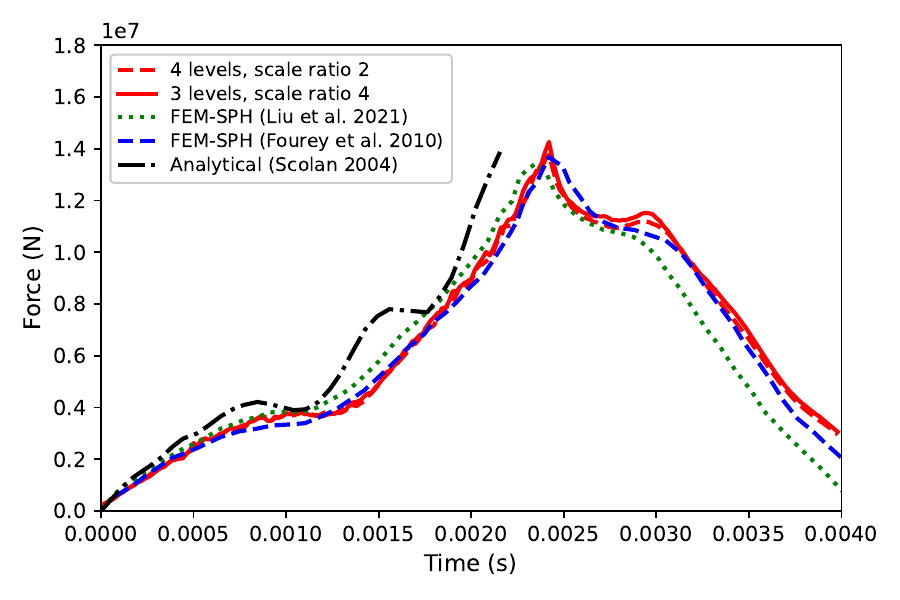}}
	
	\caption{Elastic structure impact on the free surface: wedge deflection at location C and vertical force applied on the wedge from simulation (i) and (ii), and comparisons with the results by Long et al. \cite{long2021coupling} with FEM-SPH, Fourey et al. \cite{fourey2010violent} with FEM-SPH and analytical solution by Scolan \cite{scolan2004hydroelastic}.}
	\label{wedge_defl_verf}
\end{figure*}

\begin{figure*}[htbp]
	\centering
	
	\subfigure[Probe A]{
		\includegraphics[width=0.47\linewidth]{ 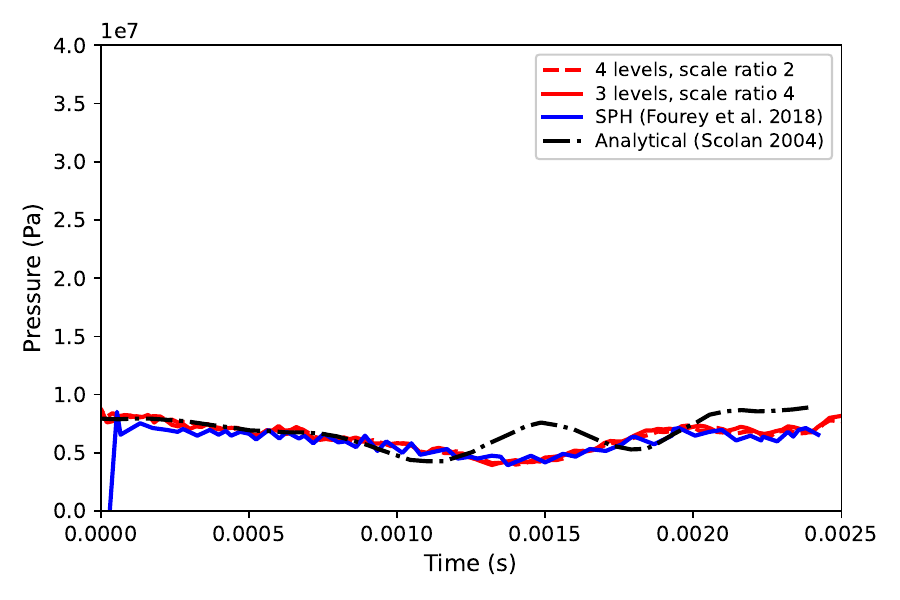}}
	\subfigure[Probe B]{
		\includegraphics[width=0.47\linewidth]{ 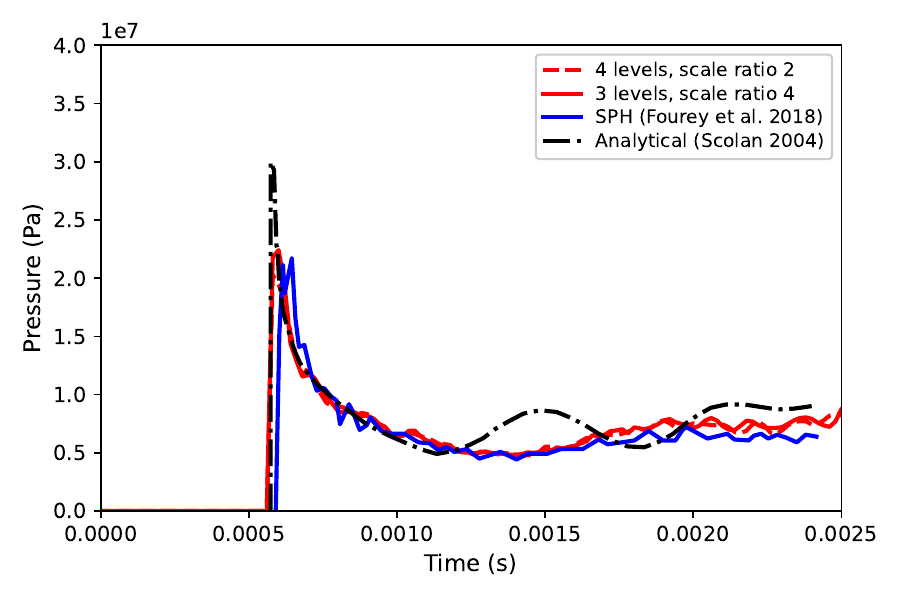}}
	
	\subfigure[Probe C]{
		\includegraphics[width=0.47\linewidth]{ 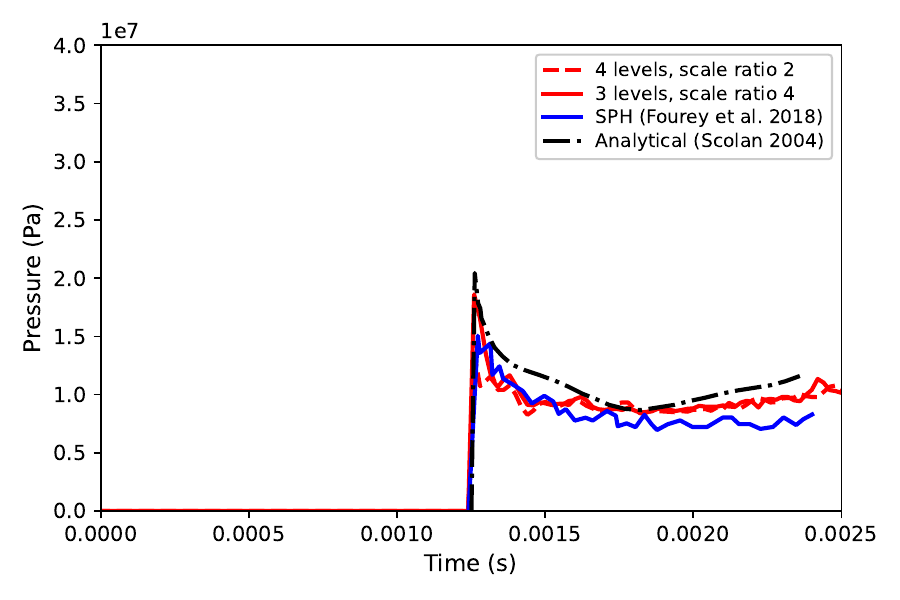}}
	\subfigure[Probe D]{
		\includegraphics[width=0.47\linewidth]{ 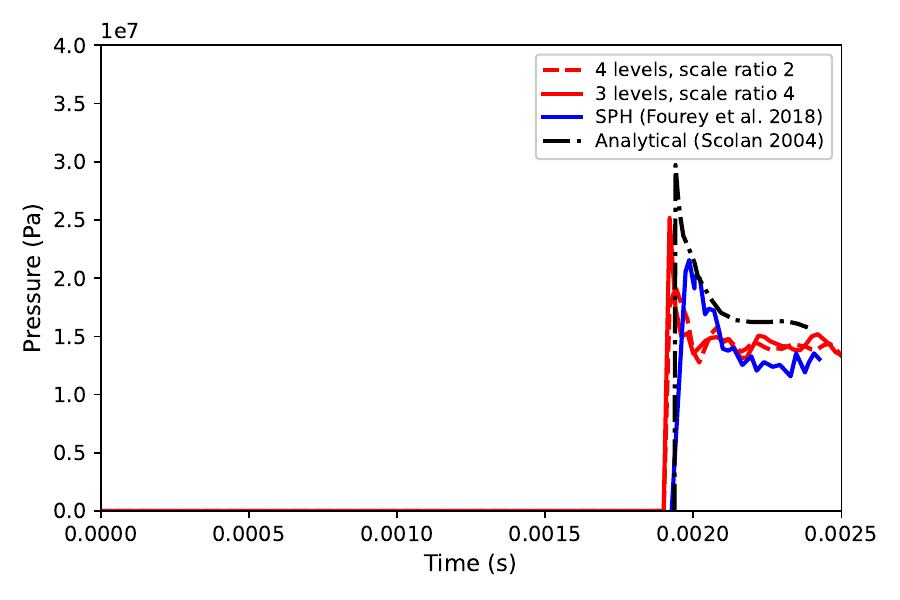}}
	\caption{Elastic structure impact on the free surface: pressure history at 4 probe locations from simulation (i) and (ii), and comparisons with the results by Oger et al. \cite{oger2009simulations} and analytical solution by Scolan \cite{scolan2004hydroelastic}. }	
	\label{wedge_P}
\end{figure*}

Fig. \ref{wedge_snap} shows the results of the two simulations (i) and (ii). The pressure distribution between particles of different resolutions is shown to be highly smooth. The deflection history at probe location A is plotted in Fig. \ref{wedge_defl_verf}(a). In this figure, the result using the refinement scale ratio 2 almost coincides with that using the scale ratio 4. The present results agree very well with the semi-analytical solution by Scolan \cite{scolan2004hydroelastic} and the results solved by the method coupling SPH with FEM by Fourey et al. \cite{fourey2010violent}, and are closer to the FEM-SPH results by Long et al. \cite{long2021coupling}. Regarding the vertical force applied on the wedge in Fig. \ref{wedge_defl_verf}(b), the results of  simulations (i) and (ii) agree with each other quite well, and the present results also have a good agreement with the results by Fourey et al. \cite{fourey2010violent}, Long et al. \cite{long2021coupling} and the semi-analytical solution \cite{scolan2004hydroelastic}. Fig. \ref{wedge_P} shows the pressure history at 4 probe locations. Likewise, the results using the refinement scale ratio 2 and 4 agree quite well with each other and also have a reasonable agreement with the results by Oger et al. \cite{oger2009simulations} and the semi-analytical solution by Scolan \cite{scolan2004hydroelastic}. This case further demonstrates that the present APR method with a large refinement scale ratio is reliable. 

\begin{figure}[htbp]
	\centering
	\includegraphics[width=0.9\textwidth]{ 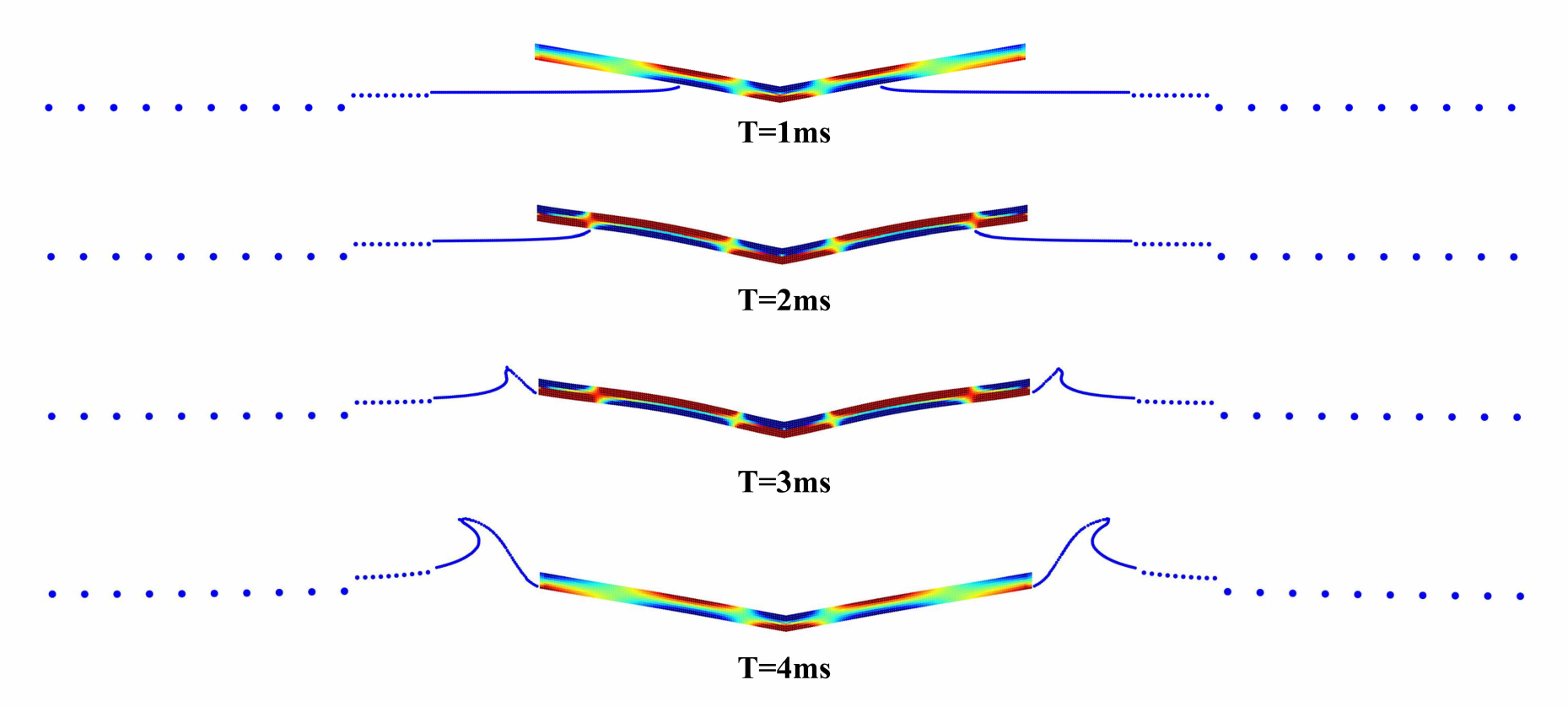}
	\caption{The free-surface particles (blue color) are identified by the new free-surface detection method in simulation (ii) with 3 resolution levels and scale ratio 4.}
	\label{wedge_surf}
\end{figure}
\subsection{Fishlike swimming problem}
\begin{figure}[htbp]
	\centering
	\includegraphics[width=0.7\textwidth]{ 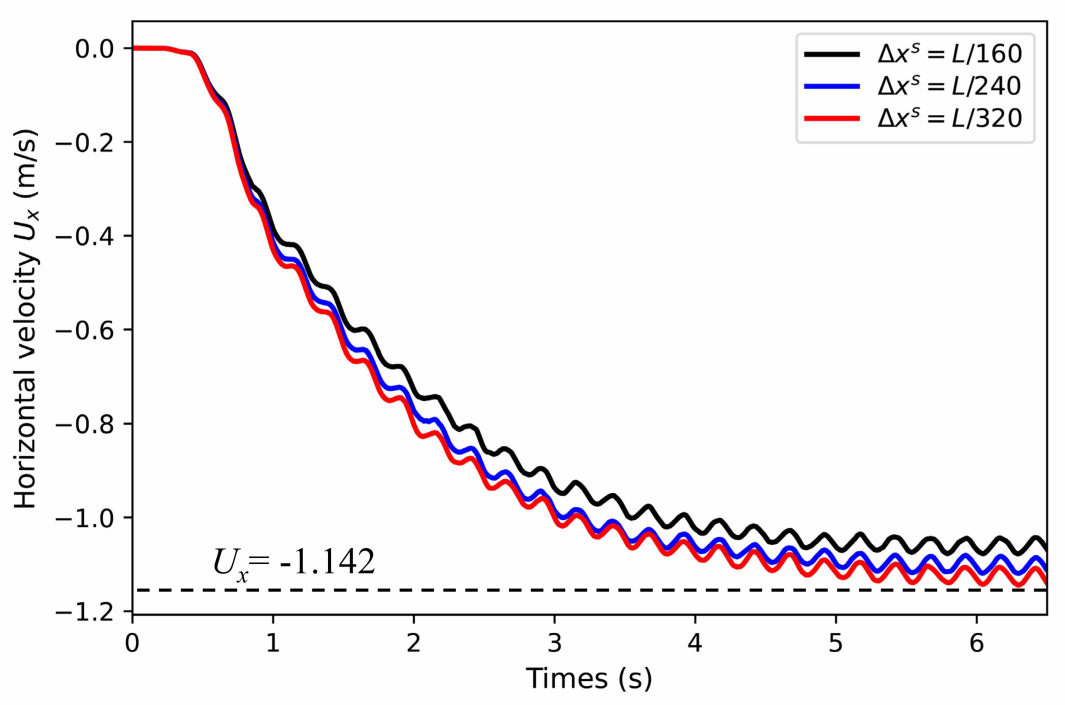}
	\caption{Horizontal velocity of the swimming fish under three different particle resolutions, with the rotation and the vertical motion of the fish constrained. }
	\label{fish_convg_plt}
\end{figure}

To mimic the shape of a fish, a NACA0012 foil is employed similar to Deng et al. \cite{deng2007hydrodynamic}, Dong and Lu \cite{dong2007characteristics} and Yang et al. \cite{yan2008two}. The stable undulation motion of the fish centerline is described by a wave function as
\begin{equation}
	y( x,t)=A(x)r(t){\cos} (kx-\dfrac{2\pi}{T_0}  t) ,
	\label{wave}
\end{equation}
where $ k=2\pi/\lambda $ is the wave number, and $ \lambda $ is the wave length. $ T_0 $ is the undulation cycle. $ A(x) $ is the undulation amplitude at different locations, expressed as
\begin{equation}
	A(x)= L \left[ a_0+ a_1 \frac{x}{L} + a_2 (\frac{x}{L})^2\right] ,
	\label{ampli}
\end{equation}
and $ r(t) $ is an increasing ramp function \cite{yan2008two} given as
\begin{equation}
	r(t)=  
	\begin{cases}
		\frac{t}{t_0}-\frac{1}{2\pi}{\sin} \left( {2\pi \frac{t}{t_0} }\right), & 0 \leq t \leq t_0,\\
		1, &  t > t_0.
	\end{cases}
	\label{ramp}
\end{equation}
Here $ L $, $ A $ and $ t_0 $ are the body length of the fish, the amplitude constant and the ramp time. According to the experimental data \cite{videler1993fish} and the simulation setup by Sun et al. \cite{sun2018numerical}, the three coefficients are determined as $ a_0= 0.02$, $ a_1= -0.0825$ and $ a_2= 0.1625$. The body length $ L $ is set as $ L=1 \text{ }\rm m $, and the wave length is $ \lambda=L $. The undulation cycle is $ T_0=0.5 \text{ }\rm s $ and ramp time $ t_0=T_0 $. The density of the fish is set equal to the fluid density $ 1 \text{ }\rm{kg/m^3} $, and the kinematic viscosity of the fluid is $ \nu=2 \times 10^{-4} \text{ }\rm {Pa \cdot s} $, which leads to an approximate Reynolds number of $ Re=5000 $. The artificial sound speed is set as $ c_0=20 \ \rm{m/s} $. In this case, 3 resolution levels with the refinement scale ratio 4 are deployed around the fish. The resolution for the fish body is $\Delta x^s=L/160 $. The curving boundary of the fishlike foil will impair the simulation accuracy in this situation, hence an isotropic particle distribution in the fish body is required. In this work, the isotropic particle distribution is generated with a boundary-modified Centroid Voronoi Particle (CVP) method, which is initially proposed by Fu et al. \cite{fu2017physics}, and the details are given in \ref{appdx1}. The kinematic motion of the swimming fish is described with a precise analytical solution elaborated in \ref{appdx2}. 
\begin{figure}[htbp]
	\centering
	\includegraphics[width=0.6\textwidth]{ 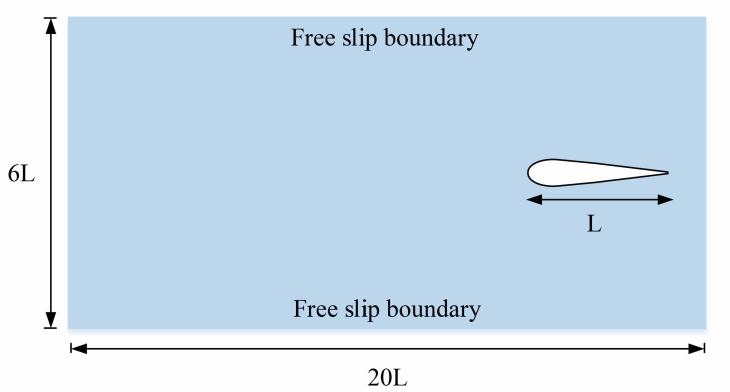}
	\caption{Schematic of fishlike swimming problem. }
	\label{schm_fish}
\end{figure}
In this case, three resolution levels with the refinement scale ratio 4 are employed. The refinement zone will be moved synchronously with the centroid of fish. Firstly, a convergence test using three different particle resolutions, i.e., $\Delta x^s=L/160$, $\Delta x^s=L/240$ and $\Delta x^s=L/320$, is investigated. In order to compare with the results from Sun et al. \cite{sun2018numerical}, the same constraints for the fish are used, i.e., the rotation and the vertical motion of the fish are constrained. The horizontal velocity statistics simulated with the three different particle resolutions are shown in Fig. \ref{fish_convg_plt}. It is seen that the cruising velocity converges to $ U_x=-1.142 $. This result is quite close to the cruising velocity $ U_x=-1.23 $, obtained by Sun et al. \cite{sun2018numerical} using resolution of  $\Delta x^s=L/400$, which demonstrates that the present method for fish swimming is reliable. 

\begin{figure}[htbp]
	\centering
	\includegraphics[width=0.9\textwidth]{ 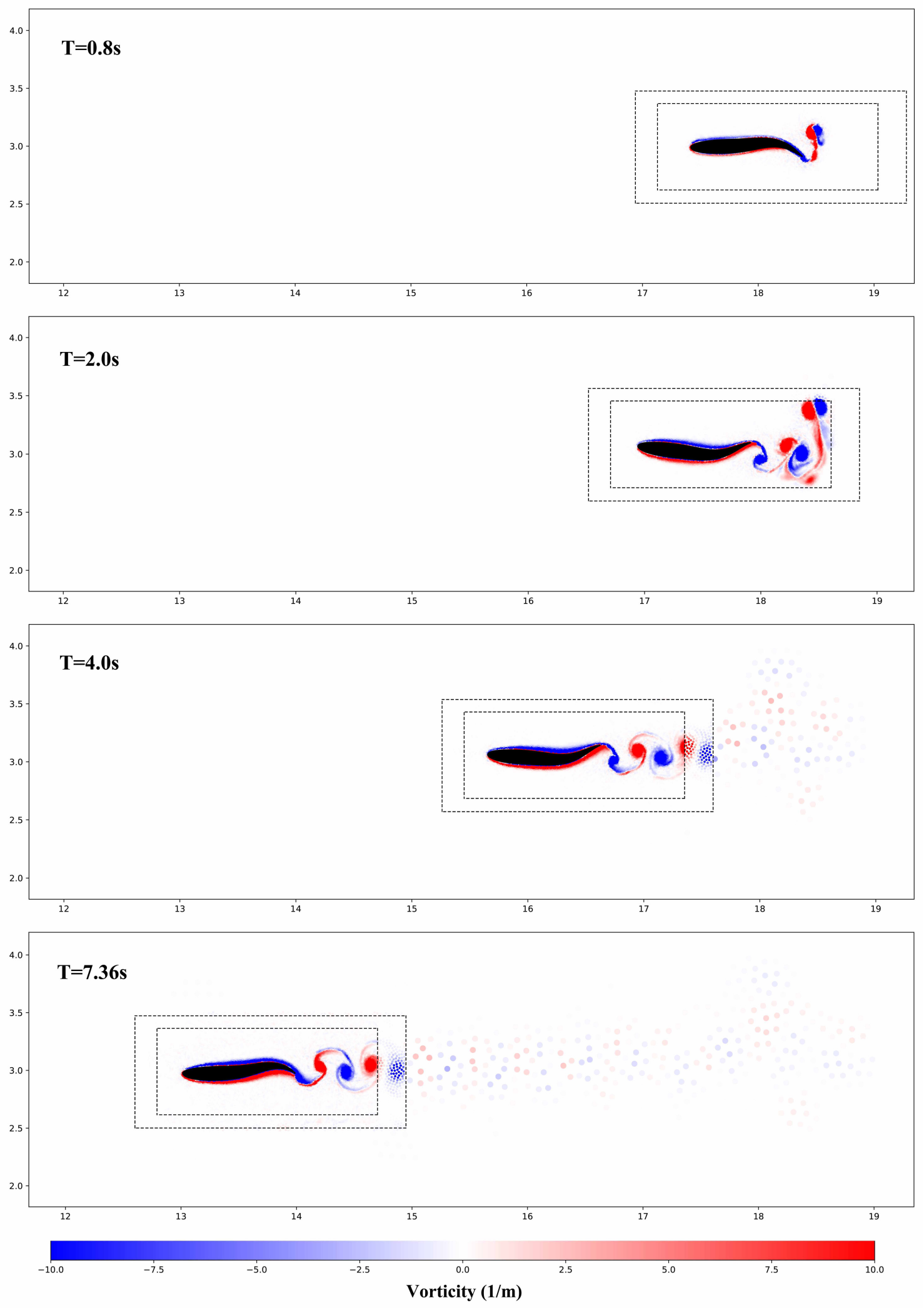}
	\caption{Fishlike swimming: vorticity fields of the fishlike swimming at different time are shown, simulated with three particle resolution levels with scale ratio 4. The refinement zone is displayed within black dashed lines.  }
	\label{fish_snap}
\end{figure}

\begin{figure}[htbp]
	\centering
	
	\subfigure[Horizontal velocity $ U_x $ and vertical velocity $ U_y $.]{
		\begin{minipage}{1\textwidth}
			\centering
			\includegraphics[width=0.7\textwidth]{ 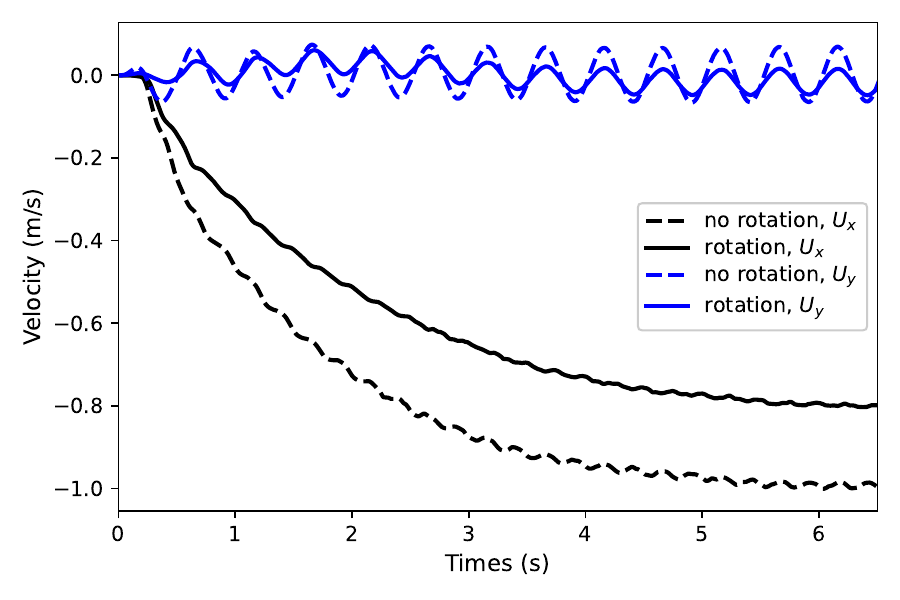}
		\end{minipage}
	}
	\subfigure[Angular velocity $ \omega $.]{
		\begin{minipage}{1\textwidth}
			\centering
			\includegraphics[width=0.7\textwidth]{ 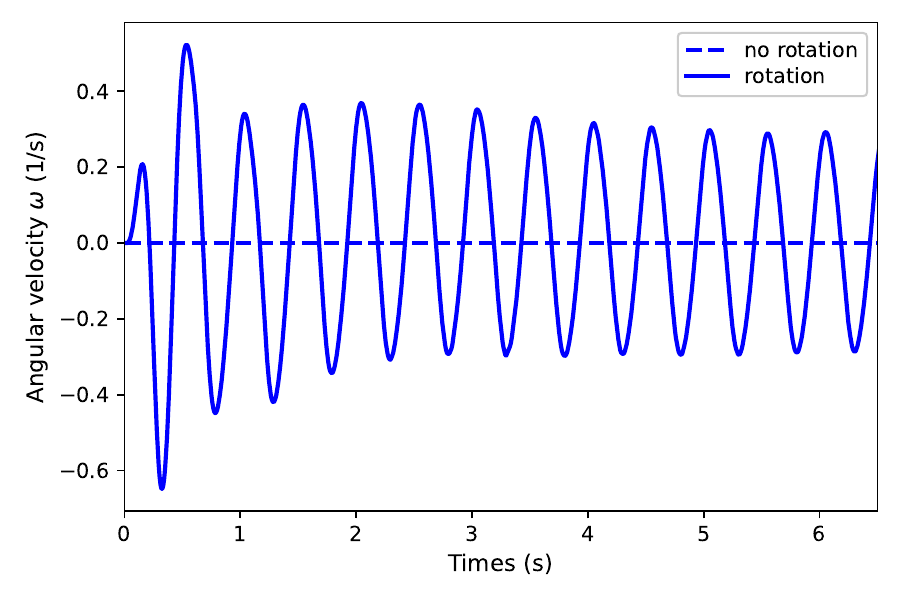}
		\end{minipage}
	}
	\caption{Horizontal velocity $ U_x $ and vertical velocity $ U_y $ (a), and angular velocity $ \omega $ (b), of the swimming fish simulated with and without allowing the rotational motion.}
	\label{vw_plt}
\end{figure}

Fig. \ref{fish_snap} shows the vorticity fields of the fishlike swimming at different time. In this case, three resolution levels with the refinement scale ratio 4 are used, and the resolution $\Delta x^s=L/160$ is used for the fish. The horizontal, vertical, and rotational motions are all allowed. The refinement zone is shown within dashed lines, and moves adaptively with the fish. The vortex structure induced by the fish motion is shown behind the fish, and it is seen that the vortex structure develops gradually to a stable state. At the finest level, the vortex is observed clearly; at the coarsest level, the vortex structure is indistinct due to the coarse resolution deployed in this area. Fig. \ref{vw_plt} shows the horizontal velocity $ U_x $, vertical velocity $ U_y $ and angular velocity $ \omega $ of the swimming fish simulated with and without rotation, denoted by scenario (i) and (ii). In Fig. \ref{vw_plt}(a), the cruising velocity of the fish is about $ U_x=-1.008 $ for scenario (i), which is slower than $ U_x=-1.142 $ simulated with the vertical motion fixed. As for scenario (ii) with the rotational motion allowed, the cruising velocity is $ U_x=-0.804 $, slower than scenario (i). Regarding the oscillatory vertical velocity, scenario (i) has a mean value equal to zero, while that of scenario (ii) diverges a little from zero. The marginal divergence may be attributed to the unstable pressure field in the initial ramp stage described by Eq. (\ref{ramp}). 

The computational efficiency using a high refinement scale ratio is investigated. In the case above, 3 resolution levels with the refinement scale ratio 4 are employed for the computation. If scale ratio 2 is employed, there will be 5 total resolution levels for achieving the same finest resolution. The computational time for the two setups is compared in Table \ref{tb:fish}. It reduces over $ 30\% $ computational time when using scale ratio 4 instead of scale ratio 2.  If the uniform resolution is used, the computational time is over 1000 s, which is prohibitively expensive compared to that using 3 resolution levels with scale ratio 4. Overall, this case further demonstrates the high performance of the present APR method. 
\begin{table}
	\renewcommand{\arraystretch}{1.3}
	\caption{Comparison of the average computational time for each 0.008s using different scale ratios and uniform resolution.} 
	\begin{center}
		%%% Some packages, such as MDW tools, offer better commands for making tables
		%%% than the plain LaTeX2e tabular which is used here.
		\begin{tabular}{p{3cm} p{5cm} p{5cm}p{2cm}}
			\toprule [1.2 pt]
			Resolution & 5 levels, scale ratio 2 & 3 levels, scale ratio 4 & Uniform\\
			\hline
			$\Delta x^s=L/160$ & 25 s & 16 s & \textgreater 1000 s \\	
			$\Delta x^s=L/240$ & 74 s & 51 s & \textgreater1000 s \\	
			\bottomrule [1.2 pt]
		\end{tabular}
	\end{center}
	\label{tb:fish}
\end{table}
\section{Concluding remarks}

In this study, the multi-resolution APR method for SPH with a large refinement scale ratio, the regularized transition sub-zone, and a new free-surface  detection algorithm is devised. The present method is applied in simulating a set of complex FSI problems, where the one-way FSI coupling strategy is implemented. Considering the practical circumstance, the multiphase model is also incorporated and used to simulate the complex FSI problem based on the present improved multi-resolution method. A set of benchmark cases, i.e., water impact on an elastic thin beam, water impact on an elastic thin beam with the two-phase model, elastic structure impact on the free surface, and fishlike swimming problem, has been studied with the present method. It turns out that, (a) the improved APR using a large scale ratio of 4 agrees very well with that using scale ratio of 2; (b) the improved APR with a large scale ratio of 4 can significantly reduce computational costs compared to that with scale ratio of 2; (c) the deployed regularized transition sub-zone ensures a smooth pressure field between the non-refinement zone and the refinement zone; (d) the proposed new free-surface detection method based on the Voronoi diagram can identify the particles at the free surface more accurately and efficiently than the conventional method proposed by Marrone et al. [27]. 
An interesting fish swimming problem is simulated in the end to further assess the performance of the improved multi-resolution framework, and a set of new algorithms, including the isotropic particle generation and the precise kinematics of fish swimming, is applied for this problem. The fishlike swimming problem demonstrates that the present method is promising for simulating complex FSI problems in bio-applications and practical engineering problems. Overall, the improved APR method with a large refinement scale ratio and the new free-surface detection method shows great potential for simulating complex FSI problems efficiently and accurately.

\section*{Acknowledgment}
% optional entry into table of contents (if used)

This work was supported by the Research Grants Council (RGC) of the Government of Hong Kong Special Administrative Region (HKSAR) with RGC/GRF Project (No. 16206321) and the fund from Shenzhen Municipal Central Government Guides Local Science and Technology Development Special Funds Funded Projects (No. 2021Szvup138). Lin Fu also acknowledges Zhe Ji, who has been working with him to develop the basic SPH platform over the past several years.

\appendix
\setcounter{figure}{0}
\section{Isotropic particle generation of fishlike foil}
\label{appdx1}
\begin{figure}[htbp]
	\centering
	\includegraphics[width=0.5\textwidth]{ 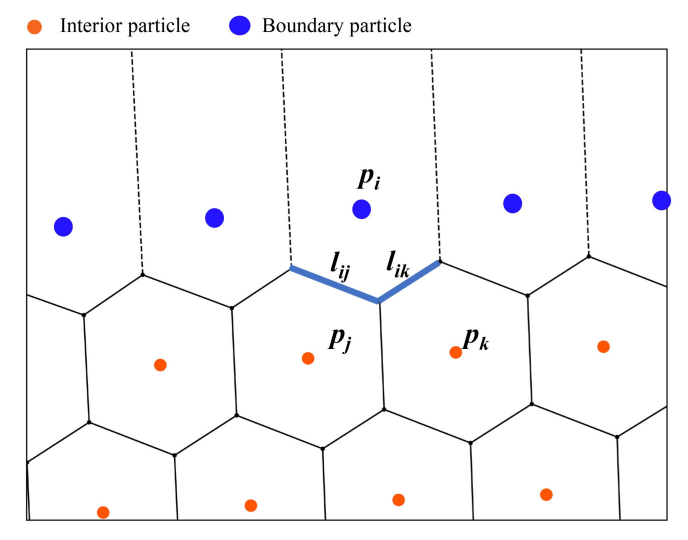}
	\caption{Schematic of the pressure assignment of the boundary particles in the modified CVP method: blue and orange particles are interior and boundary particles, respectively, and the blue bold lines are common edges of two neighboring cells. }
	\label{fish_cell}
\end{figure}

In the boundary-modified CVP method, evenly distributed particles with equal spacing $ h_e $ are generated on the boundary curve firstly with the mesh-generation module in an extra software, e.g., Abaqus. Secondly, a structured particle array is generated within the boundary curve of the fish body. Then, the CVP method \cite{fu2017physics}\cite{fu2019optimal} will be used to render these particles isotropic. The artificial pressure of each non-boundary Voronoi cell is defined as $ p_i=1/V_i $, where $ V_i $ is the volume of the Voronoi cell. For the boundary Voronoi cell, the pressure is derived by calculating the mean value from its neighboring non-boundary Voronoi cells, written as 
\begin{equation}
	p_i = \frac{\sum_{j}p_j l_{j}}{\sum_{j} l_{j}},
\end{equation}
where $ p_j $ is the pressure of neighboring Voronoi cell $ j $, and $ l_j $ is the common edge of cell $ i $ and $ j $, as shown in Fig. \ref{fish_cell}.

The acceleration of particle $ \boldsymbol{a}_{i} $ is governed by
\begin{equation}
	\boldsymbol{a}_{i} = -\int_{\Omega_i} \nabla pdV =  -\int_{\partial\Omega_i} pdS,
\end{equation}
where $ \Omega_i $ and $ \partial\Omega_i $ denote the volume and surface of Voronoi cell. The positions of particles are updated with a two-step process. In the first step, the particles are updated to $ \boldsymbol{x}_{i}^* $ by
\begin{equation}
	\boldsymbol{x}_{i}^* = \boldsymbol{x}_{i}^n + \gamma \frac{1}{2} \boldsymbol{a}_{i}^n {\Delta t}^2 ,
\end{equation}
where the time step is $ \Delta t = \min \left( 0.25\sqrt{\frac{h_e}{\left\|\boldsymbol{a}_{i} \right\| } } \right)  $ and $ \gamma=0.8 $. In the next step, the particles are updated to the new position $ \boldsymbol{x}_{i}^{n+1} $ following
\begin{equation}
	\boldsymbol{x}_{i}^{n+1} = \boldsymbol{x}_{i}^* + (1-\gamma) {\Delta \tau}(\boldsymbol{z}_{i}^n -  \boldsymbol{x}_{i}^* )  ,
\end{equation}
where $ \boldsymbol{z}_{i} $ is the centroid of Voronoi cell $ i $, and the time step is defined as
\begin{equation}
	\Delta \tau ={\min}({\min} (h_e/32, \left\| \boldsymbol{z}_{i}^n -\boldsymbol{x}_{i}^n \right\| )/\left\| \boldsymbol{z}_{i}^n -\boldsymbol{x}_{i}^n \right\| ).
\end{equation}

\begin{figure}[htbp]
	\centering
	\includegraphics[width=1\textwidth]{ 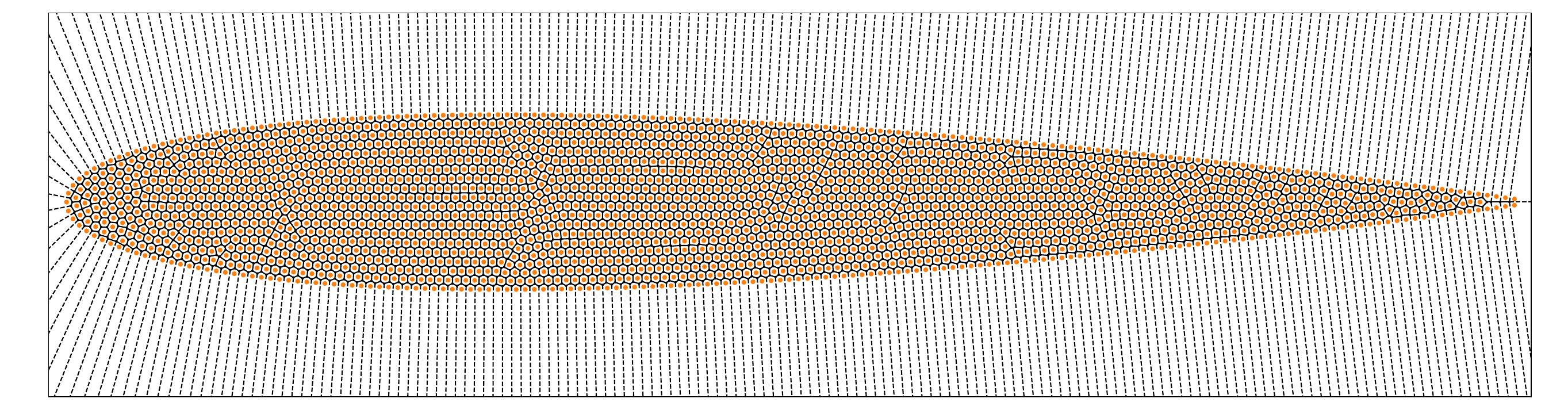}
	\caption{Isotropic particle distribution generated by the modified CVP method: the particles (Voronoi cell generators) in orange color comprise the fish body, and the lines are Voronoi cell edges. }
	\label{fish_particle}
\end{figure}

The above procedure is iterated until the convergence criterion is met. Here, we introduce the error $ e=\left\| \boldsymbol{x}_{i}^{n+1}-\boldsymbol{x}_{i}^{n}\right\|/\left\| \boldsymbol{x}_{i}^{1}-\boldsymbol{x}_{i}^{0}\right\|  $, and if $ e\textless 0.05 $, the particle distribution is treated as isotropic and the iteration is terminated. Fig. \ref{fish_particle} shows the isotropic particle distribution generated by the modified CVP method, and the particle distribution is very smooth. The modified CVP method does not need ghost particles at the geometry boundary as required by the standard CVP method and other particle regularization method. 
\setcounter{figure}{0}
\section{ Kinematics of fishlike swimming problem}
\label{appdx2}
\begin{figure}[htbp]
	\centering
	\includegraphics[width=0.6\textwidth]{ 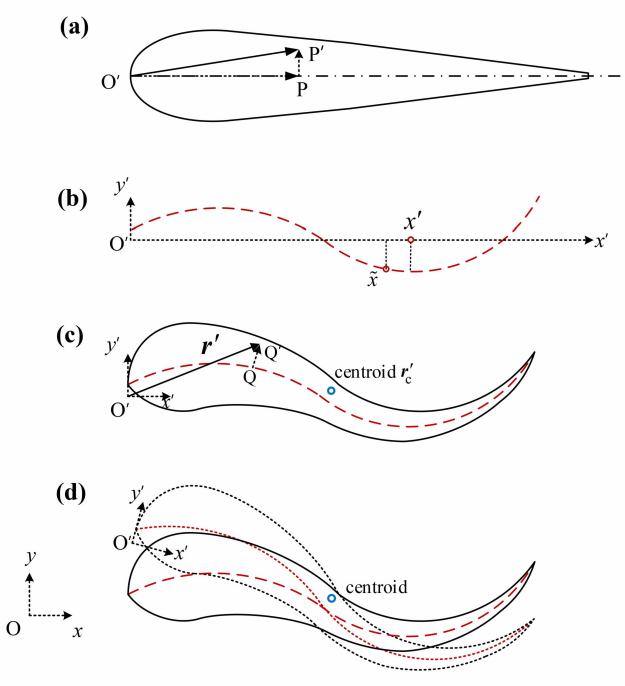}
	\caption{Schematic of fish kinematics: (a) in the fish body, $ \overline {{\rm P}{\rm P}^{\prime}}  $ is perpendicular to the fish centerline; (b) the fish centerline (red dashed line) is deformed, with location $ x $ shifting to $ \tilde{x} $; (c) deformed configuration of the fish body; (d)  rotation of the fish body with regard to the centroid. }
	\label{fish_kine}
\end{figure}

Fig. \ref{posi_fish}(a) shows the un-deformed fish body with length $L$. The motion of the fish centerline is prescribed by a function $ y^{\prime}(x^{\prime},t) $ which is usually a wave function obtained by experimental observation, where $x^{\prime}$ is   in the local coordinate system $ ({\rm O^{\prime}}, x^{\prime},y^{\prime}) $. When the centerline is deformed, the position $ x^{\prime} $ at the centerline will shift to  $ \tilde{x} $ in the local coordinate system, as shown in Fig. \ref{posi_fish}(b), which lead to the deformed centerline $ y^{\prime}(\tilde{x},t) $. Given that the fish length is not extensible,  we have
\begin{equation}
	\int_{0}^{\tilde{x}(x^{\prime})} \sqrt{\left( {\dfrac{dy^{\prime}}{d\xi}}\right)  ^2 +1} d\xi=x^{\prime} .
	\label{nonexten}
\end{equation}
Taking the derivative of Eq. (\ref{nonexten}) and considering	$ \frac{dy^{\prime}}{d\tilde{x}}=\frac{dy^{\prime}}{dx^{\prime}}/  \frac{d\tilde{x}}{dx^{\prime}} $, we have
\begin{equation}
	\dfrac{d\tilde{x}}{dx^{\prime}}=\sqrt{ 1- \left( \dfrac{dy^{\prime}}{dx^{\prime}} \right)^2 } .
	\label{dx1_dx}
\end{equation}
For each position $ {\rm P}^{\prime} $ in the un-deformed fish body, it has a corresponding position $ {\rm P} $ at the centerline, with line $ \overline {{\rm P}{\rm P}^{\prime}}  $ perpendicular to the centerline. It is assumed the line $ \overline {{\rm P}{\rm P}^{\prime}}  $ is always perpendicular to the centerline in the deformation process, and the length of $ \overline {{\rm P}{\rm P}^{\prime}}  $ is not varying. $ \overline {{\rm P}{\rm P}^{\prime}}  $ is correspondent to $ \overline {{\rm Q}{\rm Q}^{\prime}}  $ in the deformed configuration shown in Fig. \ref{fish_kine}(c). The length of line $ \overline {{\rm P}{\rm P}^{\prime}}  $ is denoted by $ l(x^{\prime}) $, which can be positive or negative depending on whether it is at the upper or lower of the centerline. With the above assumptions, the position $ \boldsymbol{r}^{\prime} $ in the fish body with regard to the local coordinate system $ ({\rm O^{\prime}}, x^{\prime},y^{\prime}) $ can be presented as
\begin{equation}
	\left\{
	\begin{array}{l}
		\begin{aligned}
			\vspace{1ex}
			&r_{x}^{\prime} = \tilde{x} - \dfrac{{dy^{\prime}}/{d\tilde{x}}}{\sqrt{\left( {dy^{\prime}}/{d\tilde{x}}\right)  ^2 +1}} l(x^{\prime}) =\tilde{x} - \dfrac{dy^{\prime}}{dx^{\prime}}l(x^{\prime}),\\
			&r_{y}^{\prime} = y^{\prime} + \dfrac{1}{\sqrt{\left( {dy^{\prime}}/{d\tilde{x}}\right)  ^2 +1}} l(x^{\prime})=y^{\prime} + \sqrt{1-\left( \dfrac{dy^{\prime}}{dx^{\prime}} \right) ^2} l(x^{\prime}) = y^{\prime} + \dfrac{d\tilde{x}}{dx^{\prime}} l(x^{\prime}).\\
		\end{aligned}
	\end{array}
	\right.
	\label{motion1}
\end{equation}
Taking the 1st-order and 2nd-order derivatives of the position with respect to time, we have the velocity vector
\begin{equation}
	\left\{
	\begin{array}{l}
		\begin{aligned}
			\vspace{1ex}
			&{\dot r}_{x}^{\prime} =\dot {\tilde{x}} - \dfrac{d \dot y^{\prime}}{dx^{\prime}}l(x^{\prime}),\\
			&{\dot r}_{y}^{\prime} = \dot y^{\prime} + \dfrac{d \dot{\tilde{x}}}{dx^{\prime}}l(x^{\prime}),\\
		\end{aligned}
	\end{array}
	\right.
\end{equation}
and acceleration vector
\begin{equation}
	\left\{
	\begin{array}{l}
		\begin{aligned}
			\vspace{1ex}
			&{\ddot r}_{x}^{\prime} =\ddot {\tilde{x}} - \dfrac{d \ddot y^{\prime}}{dx^{\prime}}l(x^{\prime}),\\
			&{\ddot r}_{y}^{\prime} = \ddot y^{\prime} + \dfrac{d \ddot{\tilde{x}}}{dx^{\prime}} l(x^{\prime}).\\
		\end{aligned}
	\end{array}
	\right.
	\label{motion2}
\end{equation}
Herein, $ {\tilde{x}} $, $ \dot {\tilde{x}} $ and $ \ddot {\tilde{x}} $ can be solved by numerical integration method with boundary condition $ \tilde{x}(0)= \dot {\tilde{x}}(0) =\ddot {\tilde{x}}(0)= 0 $, after having derived $ \frac{d{\tilde{x}}}{dx^{\prime}} $, $ \frac{d\dot {\tilde{x}}}{dx^{\prime}} $ and $ \frac{d\ddot {\tilde{x}}}{dx^{\prime}} $ from Eq. (\ref{dx1_dx}). $\frac{d \dot y^{\prime}}{dx^{\prime}}$, $\frac{d \dot{\tilde{x}}}{dx^{\prime}}$, $\frac{d \ddot y^{\prime}}{dx^{\prime}}$ and $\frac{d \ddot{\tilde{x}}}{dx^{\prime}}$ are easily derived by taking the derivatives with respect to time for $\frac{dy^{\prime}}{dx^{\prime}}$ 
and Eq. (\ref{dx1_dx}). Eqs. (\ref{motion1})$ \sim $(\ref{motion2}) provide an analytical solution for the fish deformation.

The position $ \boldsymbol{r} $ in the inertial coordinate system $ ({\rm{O}}, x,y) $  can be expressed as
\begin{equation}
	\boldsymbol{r}=\boldsymbol{r}_c+\boldsymbol{\rm R}(\theta) \boldsymbol{\bar{r}}^{\prime}=\boldsymbol{r}_c+\boldsymbol{\rm R}(\theta)(\boldsymbol{r}^{\prime} -  \boldsymbol{r}_c^{\prime}),
	\label{posi_fish}
\end{equation}
where $ \boldsymbol{r}_c $ is the fish centroid in the inertial coordinate system $ ({\rm{O}}, x,y) $, and $ \boldsymbol{r}^{\prime} $ and $ \boldsymbol{r}_c^{\prime} $ are the corresponding positions  in the local coordinate system $ ({\rm O^{\prime}}, x^{\prime},y^{\prime}) $ of $ \boldsymbol{r} $ and $ \boldsymbol{r}_c $, as shown in Fig. \ref{fish_kine} (c).
$\boldsymbol{\rm R}(\theta)  $ is the rotation matrix, written as
\begin{equation}
	\boldsymbol{\rm R}(\theta)=\begin{pmatrix}
		{\cos}\theta & -{\sin}\theta \\
		{\sin}\theta & {\cos}\theta
	\end{pmatrix}.
\end{equation}
The matrix is used to rotate the position in the local coordinate system $ ({\rm O^{\prime}}, x^{\prime},y^{\prime}) $  to the position in the inertial coordinate system $ ({\rm{O}}, x,y) $ with respect to the centroid $ \boldsymbol{r}_c $, as shown in Fig. \ref{fish_kine}(d).
With Eq. (\ref{posi_fish}), the velocity of the point within the fish, i.e., the derivative of a position vector with respect to time is obtained as
\begin{equation}
	\boldsymbol{\dot r}=\boldsymbol{\dot r}_c+\boldsymbol{\rm R} (\theta) \boldsymbol{\dot {\bar {r}}} ^{\prime} + \boldsymbol{\rm R}(\theta)\boldsymbol{\rm R}(\frac{\pi}{2}) \dot \theta \boldsymbol{ {\bar {r}}} ^{\prime} .
	\label{velo_fish}
\end{equation}
Likewise, the acceleration of the point within the fish is derived by solving the 2nd-order derivative of a position vector with respect to time, written as
\begin{equation}
	\boldsymbol{\ddot r}=\boldsymbol{\ddot r}_c+\boldsymbol{\rm R}(\theta) \boldsymbol{\ddot {\bar {r}}} ^{\prime}+ \boldsymbol{\rm R}(\theta)\boldsymbol{\rm R}(\pi) {\dot \theta}^2 \boldsymbol{ {\bar {r}}} ^{\prime}  + \boldsymbol{\rm R}(\theta)\boldsymbol{\rm R}(\frac{\pi}{2}) \ddot \theta \boldsymbol{ {\bar {r}}} ^{\prime}  + 2\boldsymbol{\rm R}(\theta)\boldsymbol{\rm R}(\frac{\pi}{2}) \dot \theta \boldsymbol{\dot {\bar {r}}} ^{\prime} ,
	\label{acce_fish}
\end{equation}
where $ \dot \theta $ and $ \ddot \theta $ are the angular velocity and angular acceleration of the fish. The centroid acceleration and angular acceleration are calculated with
\begin{equation}
	\left\{
	\begin{array}{l}
		\begin{aligned}
			\vspace{1ex}
			&\dfrac{ d(m\boldsymbol{\dot r}_c)}{dt} =\boldsymbol{F}_e,\\
			&\dfrac{ d(I_c{\dot \theta})}{dt} =\boldsymbol{M}_{e}^c.\\
		\end{aligned}
	\end{array}
	\right.
\end{equation}
Here, $ m $ and $ I_c $ are the mass and moment of inertia with regard to the centroid, respectively. The calculations of external force $ \boldsymbol{F}_e $ and external torque $ \boldsymbol{M}_{e}^c $ applied by the fluid are referred to \cite{bouscasse2013nonlinear} for more details. $ \boldsymbol{r}_c^{\prime} $ , $ \boldsymbol{\dot r}_c^{\prime} $ and $ \boldsymbol{\ddot r}_c^{\prime} $ in Eqs. (\ref{posi_fish}), (\ref{velo_fish}) and (\ref{acce_fish}) can be written as
\begin{equation}
\label{eq:r_c}
	\left\{
	\begin{array}{l}
		\begin{aligned}
		\vspace{1ex}
        	&\boldsymbol{r}_c^{\prime}=\dfrac{\int _{0}^{L} l(x^{\prime}) (\tilde{x},\ y(x^{\prime},t))^T dx^{\prime}}{\int _{0}^{L} l(x^{\prime}) dx^{\prime}},\\
        	&\boldsymbol{\dot r}_c^{\prime}=\dfrac{\int _{0}^{L} l(x^{\prime}) (\dot {\tilde{x}},\ \dot y(x^{\prime},t))^T dx^{\prime}}{\int _{0}^{L} l(x^{\prime}) dx^{\prime}},\\
        	&\boldsymbol{\ddot r}_c^{\prime}=\dfrac{\int _{0}^{L} l(x^{\prime}) (\ddot {\tilde{x}},\ \ddot y(x^{\prime},t))^T dx^{\prime}}{\int _{0}^{L} l(x^{\prime}) dx^{\prime}}.
		\end{aligned}
	\end{array}
	\right.
\end{equation}
Given $ {\tilde{x}} $, $ \dot {\tilde{x}} $ and $\ddot {\tilde{x}}$ are derived, Eq. (\ref{eq:r_c}) can be calculated by a simple numerical integration method.

\setcounter{figure}{0}
\section{\textcolor{black}{Hydrostatic problem}}
\label{appdx3}
\textcolor{black}{
In order to further validate the numerical stability of simulations with the large refinement scale ratio in the improved APR method, the hydrostatic case is simulated. The water column is 5 m in width and 2.5 m in height. The property of water is the same as that in section \ref{sect:imp}. The refinement zone is deployed in the center of the water, and the scale ratio is 4 with 3 refinements levels.}

\begin{figure}[htbp]
	\centering
	
	\subfigure[Pressure field at $t=5$ s.]{
			\includegraphics[width=0.47\textwidth]{ 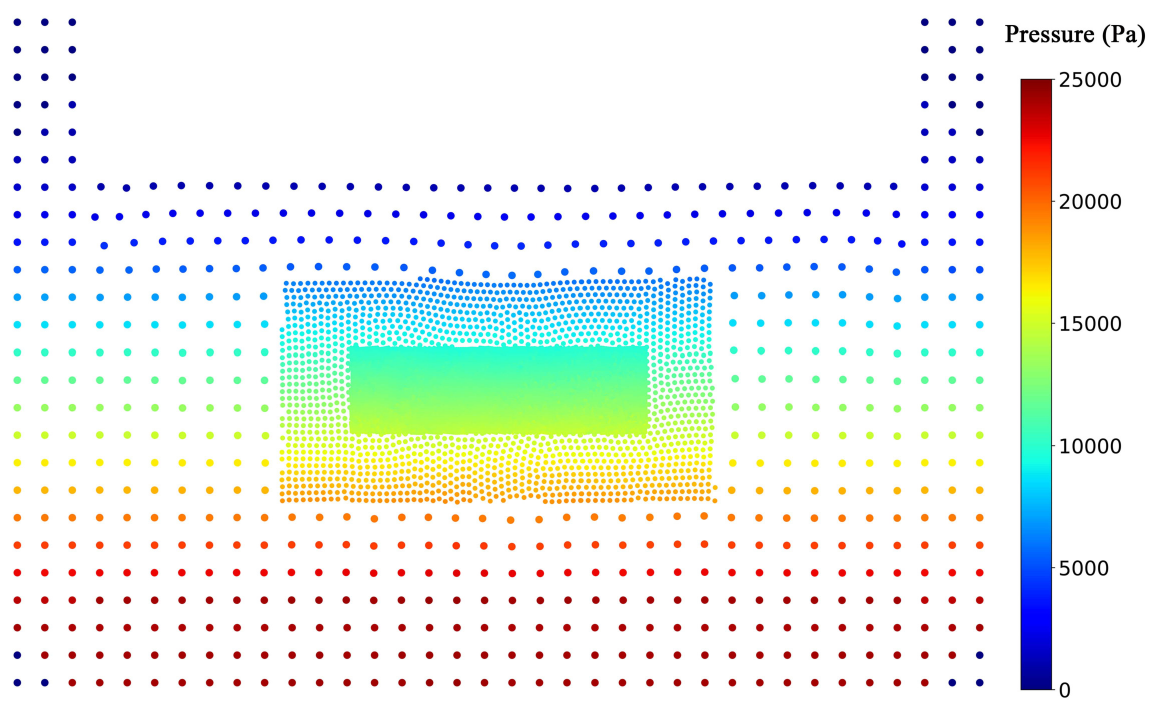}

	}
	\subfigure[Pressure history at the bottom of the water.]{
			\includegraphics[width=0.47\textwidth]{ 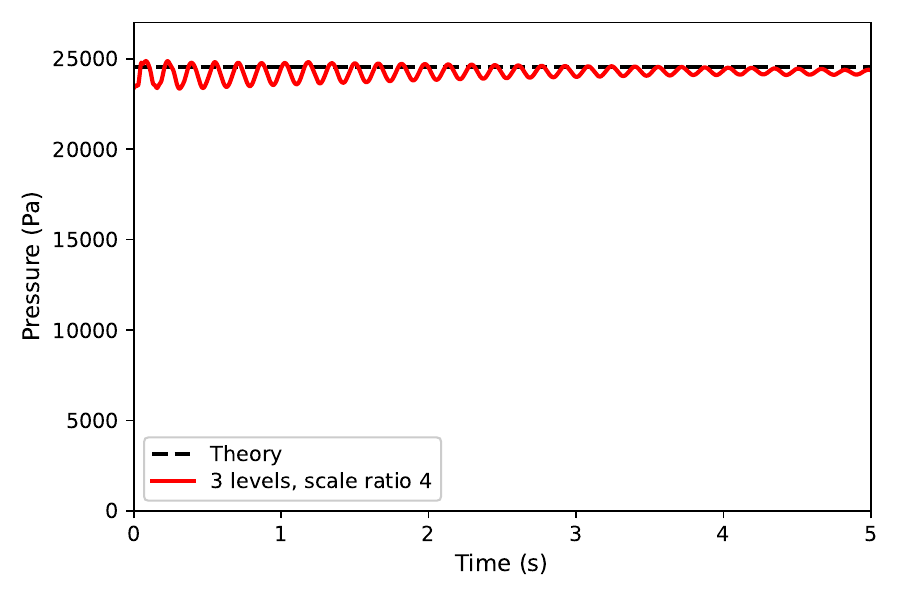}
	}
	\caption{\textcolor{black}{Hydro-static water problem: (a) the pressure field and (b) the pressure history statistics at the bottom of the water.}}
	\label{hydro}
\end{figure}
\textcolor{black}{
The pressure field at $t$=5 s is shown in Fig. \ref{hydro}(a), and it is observed that the pressure field is very smooth and stable even after a long-time simulation. The pressure history at the bottom of the water is illustrated in Fig. \ref{hydro}(b), which indicates that the predicted pressure gradually converges to the theoretical value. This case demonstrates that the improved APR method with the large refinement scale ratio of 4 is stable in the hydrostatic case. }

%% References
%%lastanswerpage


\bibliographystyle{elsarticle-num}
\scriptsize
\setlength{\bibsep}{0.5ex}
% \bibliographystyle{elsarticle-harv}
% \bibliographystyle{elsarticle-num-names}
% \bibliographystyle{model1a-num-names}
% \bibliographystyle{model1b-num-names}
% \bibliographystyle{model1c-num-names}
% \bibliographystyle{model2-names}
% \bibliographystyle{model3a-num-names}
% \bibliographystyle{model3-num-names}
% \bibliographystyle{model4-names}
% \bibliographystyle{model5-names}
% \bibliographystyle{model6-num-names}

%%\bibliography{bib.bib}

\begin{thebibliography}{10}
\expandafter\ifx\csname url\endcsname\relax
  \def\url#1{\texttt{#1}}\fi
\expandafter\ifx\csname urlprefix\endcsname\relax\def\urlprefix{URL }\fi
\expandafter\ifx\csname href\endcsname\relax
  \def\href#1#2{#2} \def\path#1{#1}\fi

\bibitem{takashi1994ale}
N.~Takashi, {ALE finite element computations of fluid-structure interaction
  problems}, Computer Methods in Applied Mechanics and Engineering 112~(1-4)
  (1994) 291--308.

\bibitem{peskin2002immersed}
C.~S. Peskin, The immersed boundary method, Acta numerica 11 (2002) 479--517.

\bibitem{sun2019study}
P.~Sun, D.~Le~Touz{\'e}, A.-M. Zhang, {Study of a complex fluid-structure
  dam-breaking benchmark problem using a multi-phase SPH method with APR},
  Engineering Analysis with Boundary Elements 104 (2019) 240--258.

\bibitem{zhang2021multi}
C.~Zhang, M.~Rezavand, X.~Hu, {A multi-resolution SPH method for
  fluid-structure interactions}, Journal of Computational Physics 429 (2021)
  110028.

\bibitem{o2021fluid}
J.~O’Connor, B.~D. Rogers, {A fluid--structure interaction model for
  free-surface flows and flexible structures using smoothed particle
  hydrodynamics on a GPU}, Journal of Fluids and Structures 104 (2021) 103312.

\bibitem{liu2013numerical}
M.~Liu, J.~Shao, H.~Li, Numerical simulation of hydro-elastic problems with
  smoothed particle hydrodynamics method, Journal of Hydrodynamics 25~(5)
  (2013) 673--682.

\bibitem{khayyer2019multi}
A.~Khayyer, N.~Tsuruta, Y.~Shimizu, H.~Gotoh, {Multi-resolution MPS for
  incompressible fluid-elastic structure interactions in ocean engineering},
  Applied Ocean Research 82 (2019) 397--414.

\bibitem{khayyer20213d}
A.~Khayyer, H.~Gotoh, Y.~Shimizu, Y.~Nishijima, {A 3D Lagrangian meshfree
  projection-based solver for hydroelastic Fluid-Structure-Interaction},
  Journal of Fluids and Structures 105 (2021) 103342.

\bibitem{zhang2019predicting}
Z.~Zhang, D.~Feng, T.~Ma, M.~Liu, Predicting the damage on a target plate
  produced by hypervelocity impact using a decoupled finite particle method,
  Engineering Analysis with Boundary Elements 98 (2019) 110--125.

\bibitem{idelsohn2008unified}
S.~R. Idelsohn, J.~Marti, A.~Limache, E.~O{\~n}ate, {Unified Lagrangian
  formulation for elastic solids and incompressible fluids: application to
  fluid--structure interaction problems via the PFEM}, Computer Methods in
  Applied Mechanics and Engineering 197~(19-20) (2008) 1762--1776.

\bibitem{colagrossi2003numerical}
A.~Colagrossi, M.~Landrini, Numerical simulation of interfacial flows by
  smoothed particle hydrodynamics, Journal of computational physics 191~(2)
  (2003) 448--475.

\bibitem{ganzenmuller2015hourglass}
G.~C. Ganzenm{\"u}ller, {An hourglass control algorithm for Lagrangian smooth
  particle hydrodynamics}, Computer Methods in Applied Mechanics and
  Engineering 286 (2015) 87--106.

\bibitem{zhu2022dynamic}
Y.~Zhu, C.~Zhang, X.~Hu, {A dynamic relaxation method with operator splitting
  and random-choice strategy for SPH}, Journal of Computational Physics 458
  (2022) 111105.

\bibitem{sun2021accurate}
P.-N. Sun, D.~Le~Touze, G.~Oger, A.-M. Zhang, An accurate fsi-sph modeling of
  challenging fluid-structure interaction problems in two and three dimensions,
  Ocean Engineering 221 (2021) 108552.

\bibitem{ji2019new}
Z.~Ji, L.~Fu, X.~Y. Hu, N.~A. Adams, {A new multi-resolution parallel framework
  for SPH}, Computer Methods in Applied Mechanics and Engineering 346 (2019)
  1156--1178.

\bibitem{ji2019lagrangian}
Z.~Ji, L.~Fu, X.~Y. Hu, N.~A. Adams, {A Lagrangian Inertial Centroidal Voronoi
  Particle method for dynamic load balancing in particle-based simulations},
  Computer Physics Communications 239 (2019) 53--63.

\bibitem{zhang2021integrative}
C.~Zhang, J.~Wang, M.~Rezavand, D.~Wu, X.~Hu, An integrative smoothed particle
  hydrodynamics method for modeling cardiac function, Computer Methods in
  Applied Mechanics and Engineering 381 (2021) 113847.

\bibitem{jacob2021arbitrary}
B.~Jacob, B.~Drawert, T.-M. Yi, L.~Petzold, {An arbitrary Lagrangian Eulerian
  smoothed particle hydrodynamics (ALE-SPH) method with a boundary volume
  fraction formulation for fluid-structure interaction}, Engineering Analysis
  with Boundary Elements 128 (2021) 274--289.

\bibitem{lai2022multiphase}
X.~Lai, S.~Li, J.~Yan, L.~Liu, A.-M. Zhang, Multiphase large-eddy simulations
  of human cough jet development and expiratory droplet dispersion, Journal of
  Fluid Mechanics 942.

\bibitem{feldman2007dynamic}
J.~Feldman, J.~Bonet, {Dynamic refinement and boundary contact forces in SPH
  with applications in fluid flow problems}, International Journal for
  Numerical Methods in Engineering 72~(3) (2007) 295--324.

\bibitem{vacondio2013variable}
R.~Vacondio, B.~Rogers, P.~K. Stansby, P.~Mignosa, J.~Feldman, {Variable
  resolution for SPH: a dynamic particle coalescing and splitting scheme},
  Computer Methods in Applied Mechanics and Engineering 256 (2013) 132--148.

\bibitem{kitsionas2002smoothed}
S.~Kitsionas, A.~Whitworth, Smoothed particle hydrodynamics with particle
  splitting, applied to self-gravitating collapse, Monthly Notices of the Royal
  Astronomical Society 330~(1) (2002) 129--136.

\bibitem{reyes2013dynamic}
Y.~Reyes~L{\'o}pez, D.~Roose, C.~Recarey~Morfa, {Dynamic particle refinement in
  SPH: application to free surface flow and non-cohesive soil simulations},
  Computational Mechanics 51~(5) (2013) 731--741.

\bibitem{barcarolo2014adaptive}
D.~A. Barcarolo, D.~Le~Touz{\'e}, G.~Oger, F.~De~Vuyst, Adaptive particle
  refinement and derefinement applied to the smoothed particle hydrodynamics
  method, Journal of Computational Physics 273 (2014) 640--657.

\bibitem{chiron2018analysis}
L.~Chiron, G.~Oger, M.~De~Leffe, D.~Le~Touz{\'e}, {Analysis and improvements of
  Adaptive Particle Refinement (APR) through CPU time, accuracy and robustness
  considerations}, Journal of Computational Physics 354 (2018) 552--575.

\bibitem{sun2019extension}
P.~Sun, A.~Colagrossi, D.~Le~Touz{\'e}, A.-M. Zhang, {Extension of the
  $\delta$-Plus-SPH model for simulating Vortex-Induced-Vibration problems},
  Journal of Fluids and Structures 90 (2019) 19--42.

\bibitem{hermange20193d}
C.~Hermange, G.~Oger, Y.~Le~Chenadec, D.~Le~Touz{\'e}, {A 3D SPH--FE coupling
  for FSI problems and its application to tire hydroplaning simulations on
  rough ground}, Computer Methods in Applied Mechanics and Engineering 355
  (2019) 558--590.

\bibitem{lyu20223d}
H.-G. Lyu, P.-N. Sun, J.-M. Miao, A.-M. Zhang, {3D multi-resolution SPH
  modeling of the water entry dynamics of free-fall lifeboats}, Ocean
  Engineering 257 (2022) 111648.

\bibitem{gao2022block}
T.~Gao, H.~Qiu, L.~Fu, {A block-based adaptive particle refinement SPH method
  for fluid--structure interaction problems}, Computer Methods in Applied
  Mechanics and Engineering 399 (2022) 115356.

\bibitem{adami2013transport}
S.~Adami, X.~Hu, N.~A. Adams, A transport-velocity formulation for smoothed
  particle hydrodynamics, Journal of Computational Physics 241 (2013) 292--307.

\bibitem{oger2016sph}
G.~Oger, S.~Marrone, D.~Le~Touz{\'e}, M.~De~Leffe, {SPH accuracy improvement
  through the combination of a quasi-Lagrangian shifting transport velocity and
  consistent ALE formalisms}, Journal of Computational Physics 313 (2016)
  76--98.

\bibitem{marrone2010fast}
S.~Marrone, A.~Colagrossi, D.~Le~Touz{\'e}, G.~Graziani, {Fast free-surface
  detection and level-set function definition in SPH solvers}, Journal of
  Computational Physics 229~(10) (2010) 3652--3663.

\bibitem{sun2017deltaplus}
P.~Sun, A.~Colagrossi, S.~Marrone, A.~Zhang, {The $\delta$plus-SPH model:
  Simple procedures for a further improvement of the SPH scheme}, Computer
  Methods in Applied Mechanics and Engineering 315 (2017) 25--49.

\bibitem{molteni2009simple}
D.~Molteni, A.~Colagrossi, {A simple procedure to improve the pressure
  evaluation in hydrodynamic context using the SPH}, Computer Physics
  Communications 180~(6) (2009) 861--872.

\bibitem{marrone2011delta}
S.~Marrone, M.~Antuono, A.~Colagrossi, G.~Colicchio, D.~Le~Touz{\'e},
  G.~Graziani, {$\delta$-SPH model for simulating violent impact flows},
  Computer Methods in Applied Mechanics and Engineering 200~(13-16) (2011)
  1526--1542.

\bibitem{antuono2021delta}
M.~Antuono, P.~Sun, S.~Marrone, A.~Colagrossi, {The $\delta$-ALE-SPH model: an
  arbitrary Lagrangian-Eulerian framework for the $\delta$-SPH model with
  particle shifting technique}, Computers \& Fluids 216 (2021) 104806.

\bibitem{xu2009accuracy}
R.~Xu, P.~Stansby, D.~Laurence, {Accuracy and stability in incompressible SPH
  (ISPH) based on the projection method and a new approach}, Journal of
  computational Physics 228~(18) (2009) 6703--6725.

\bibitem{monaghan2005smoothed}
J.~J. Monaghan, Smoothed particle hydrodynamics, Reports on progress in physics
  68~(8) (2005) 1703.

\bibitem{zhang2017generalized}
C.~Zhang, X.~Y. Hu, N.~A. Adams, A generalized transport-velocity formulation
  for smoothed particle hydrodynamics, Journal of Computational Physics 337
  (2017) 216--232.

\bibitem{he2022stable}
F.~He, H.~Zhang, C.~Huang, M.~Liu, {A stable SPH model with large CFL numbers
  for multi-phase flows with large density ratios}, Journal of Computational
  Physics (2022) 110944.

\bibitem{zhang2022efficient}
C.~Zhang, Y.~Zhu, X.~Lyu, X.~Hu, {An efficient and generalized solid boundary
  condition for SPH: Applications to multi-phase flow and fluid--structure
  interaction}, European Journal of Mechanics-B/Fluids 94 (2022) 276--292.

\bibitem{grenier2009hamiltonian}
N.~Grenier, M.~Antuono, A.~Colagrossi, D.~Le~Touz{\'e}, B.~Alessandrini, {An
  Hamiltonian interface SPH formulation for multi-fluid and free surface
  flows}, Journal of Computational Physics 228~(22) (2009) 8380--8393.

\bibitem{chen2015sph}
Z.~Chen, Z.~Zong, M.~Liu, L.~Zou, H.~Li, C.~Shu, {An SPH model for multiphase
  flows with complex interfaces and large density differences}, Journal of
  Computational Physics 283 (2015) 169--188.

\bibitem{basar2000nonlinear}
Y.~Basar, D.~Weichert, Nonlinear continuum mechanics of solids: fundamental
  mathematical and physical concepts, Springer Science \& Business Media, 2000.

\bibitem{adami2012generalized}
S.~Adami, X.~Y. Hu, N.~A. Adams, A generalized wall boundary condition for
  smoothed particle hydrodynamics, Journal of Computational Physics 231~(21)
  (2012) 7057--7075.

\bibitem{meng2022hydroelastic}
Z.-F. Meng, A.-M. Zhang, J.-L. Yan, P.-P. Wang, A.~Khayyer, {A hydroelastic
  fluid--structure interaction solver based on the Riemann-SPH method},
  Computer Methods in Applied Mechanics and Engineering 390 (2022) 114522.

\bibitem{dilts2000moving}
G.~A. Dilts, {Moving least-squares particle hydrodynamics II: conservation and
  boundaries}, International Journal for numerical methods in engineering
  48~(10) (2000) 1503--1524.

\bibitem{koshizuka1996moving}
S.~Koshizuka, Y.~Oka, Moving-particle semi-implicit method for fragmentation of
  incompressible fluid, Nuclear science and engineering 123~(3) (1996)
  421--434.

\bibitem{lee2008comparisons}
E.-S. Lee, C.~Moulinec, R.~Xu, D.~Violeau, D.~Laurence, P.~Stansby,
  {Comparisons of weakly compressible and truly incompressible algorithms for
  the SPH mesh free particle method}, Journal of computational Physics 227~(18)
  (2008) 8417--8436.

\bibitem{schaling2011boost}
B.~Sch{\"a}ling, {The boost C++ libraries}, Boris Sch{\"a}ling, 2011.

\bibitem{shepard1968two}
D.~Shepard, A two-dimensional interpolation function for irregularly-spaced
  data, in: Proceedings of the 1968 23rd ACM national conference, 1968, pp.
  517--524.

\bibitem{monaghan2000sph}
J.~J. Monaghan, {SPH without a tensile instability}, Journal of computational
  physics 159~(2) (2000) 290--311.

\bibitem{fu2019parallel}
L.~Fu, Z.~Ji, X.~Y. Hu, N.~A. Adams, Parallel fast-neighbor-searching and
  communication strategy for particle-based methods, Engineering Computations
  36 (2019) 899--929.

\bibitem{liao2015free}
K.~Liao, C.~Hu, M.~Sueyoshi, {Free surface flow impacting on an elastic
  structure: Experiment versus numerical simulation}, Applied Ocean Research 50
  (2015) 192--208.

\bibitem{marino2021numerical}
J.~M. Mari{\~n}o~Salguero, Numerical simulation of free surface flows
  interacting with flexible structures.

\bibitem{baraglia2021corotational}
F.~Baraglia, W.~Benguigui, R.~Den{\`e}fle, A corotational finite element
  approach coupled to a discrete forcing method to solve hyperelastic
  deformation induced by two-phase flow, Journal of Fluids and Structures 107
  (2021) 103403.

\bibitem{scolan2004hydroelastic}
Y.-M. Scolan, Hydroelastic behaviour of a conical shell impacting on a
  quiescent-free surface of an incompressible liquid, Journal of Sound and
  Vibration 277~(1-2) (2004) 163--203.

\bibitem{fourey2010violent}
G.~Fourey, G.~Oger, D.~Le~Touz{\'e}, B.~Alessandrini, Violent fluid-structure
  interaction simulations using a coupled sph/fem method, in: IOP conference
  series: materials science and engineering, Vol.~10, IOP Publishing, 2010, p.
  012041.

\bibitem{long2021coupling}
T.~Long, C.~Huang, D.~Hu, M.~Liu, Coupling edge-based smoothed finite element
  method with smoothed particle hydrodynamics for fluid structure interaction
  problems, Ocean Engineering 225 (2021) 108772.

\bibitem{oger2009simulations}
G.~Oger, P.-M. Guilcher, E.~Jacquin, L.~Brosset, J.-B. Deuff, D.~Le~Touz{\'e},
  Simulations of hydro-elastic impacts using a parallel sph model, in: The
  Nineteenth International Offshore and Polar Engineering Conference, OnePetro,
  2009.

\bibitem{deng2007hydrodynamic}
J.~Deng, X.-M. Shao, Z.-S. Yu, Hydrodynamic studies on two traveling wavy foils
  in tandem arrangement, Physics of fluids 19~(11) (2007) 113104.

\bibitem{dong2007characteristics}
G.-J. Dong, X.-Y. Lu, Characteristics of flow over traveling wavy foils in a
  side-by-side arrangement, Physics of fluids 19~(5) (2007) 057107.

\bibitem{yan2008two}
Y.~Yan, W.~Guan-Hao, Y.~Yong-Liang, T.~Bing-Gang, {Two-dimensional
  self-propelled fish motion in medium: An integrated method for deforming body
  dynamics and unsteady fluid dynamics}, Chinese Physics Letters 25~(2) (2008)
  597.

\bibitem{videler1993fish}
J.~J. Videler, Fish swimming, Vol.~10, Springer Science \& Business Media,
  1993.

\bibitem{sun2018numerical}
P.-N. Sun, A.~Colagrossi, A.-M. Zhang, {Numerical simulation of the
  self-propulsive motion of a fishlike swimming foil using the $\delta$+-SPH
  model}, Theoretical and Applied Mechanics Letters 8~(2) (2018) 115--125.

\bibitem{fu2017physics}
L.~Fu, X.~Y. Hu, N.~A. Adams, {A physics-motivated Centroidal Voronoi Particle
  domain decomposition method}, Journal of Computational Physics 335 (2017)
  718--735.

\bibitem{fu2019optimal}
L.~Fu, Z.~Ji, {An optimal particle setup method with Centroidal Voronoi
  Particle dynamics}, Computer Physics Communications 234 (2019) 72--92.

\bibitem{bouscasse2013nonlinear}
B.~Bouscasse, A.~Colagrossi, S.~Marrone, M.~Antuono, {Nonlinear water wave
  interaction with floating bodies in SPH}, Journal of Fluids and Structures 42
  (2013) 112--129.

\end{thebibliography}

\end{document}